\documentstyle[aps,twocolumn,eqsecnum]{revtex}
\begin{document}

\wideabs{
\title{Measuring gravitational waves from binary black hole coalescences:
II. The waves' information and its extraction, with and without templates.}

\author{\'Eanna \'E.\ Flanagan}
\address{Cornell University, Newman Laboratory, Ithaca, NY
14853-5001.}

\author{Scott A.\ Hughes}
\address{Theoretical Astrophysics, California Institute of Technology,
Pasadena, California 91125.}

\maketitle


\begin{abstract} 

We discuss the extraction of information
from detected binary black hole (BBH) coalescence gravitational wave
bursts, focusing in particular on the nonlinear {\it merger} phase of
the coalescence, which occurs after the gradual {\it inspiral} of the
bodies in the binary and before the {\it ringdown} of the system to
its final Kerr black hole state.

We report four principal results: (i) If numerical relativity
simulations have not successfully produced theoretical template
waveforms for the merger by the time that BBH waves are first detected
by LIGO/VIRGO interferometers, or if they cannot produce a set of
templates that completely covers the space of merger waveforms, then
observers can use simple band-pass filters to study the merger waves.
For BBHs of total mass $\alt 40 M_\odot$ which are detected via their
inspiral waves, we estimate that the signal-to-noise ratio from
band-pass filtering will typically be of order unity for initial and
advanced LIGO interferometers.  Thus, the merger waves should be just
visible above the noise for typical events; rare, stronger events will
be more visible, and thus more interesting.  (ii) We use Bayesian
statistics and the maximum likelihood framework to sketch out an
optimized method for extracting the merger waveform from the detector
output.  The method is based on a ``perpendicular projection'' of the
observed (noisy) signal onto an appropriate function space that
incorporates all our (possibly sketchy) prior knowledge of the
waveforms.  We argue that the best type of ``basis functions'' to use
to specify this function space is wavelets or wavelet-like functions,
and we develop the method in some detail in the language of wavelets.
In an Appendix, we sketch an extension of the method which allows one
to reconstruct the two independent polarization components of the
merger waves from the outputs of a network of several interferometers.
(iii) We propose a computational strategy for numerical relativists to
pursue, if they successfully produce computer codes for generating
merger waveforms, but if running the codes is too expensive to permit
an extensive survey of the merger parameter space.  In this case, for
LIGO/VIRGO data analysis purposes, it would be advantageous to do a
very coarse survey of the parameter space aimed at exploring several
qualitative issues and at determining the ranges of the several key
parameters which we describe.  (iv) If merger templates are available
for data analysis, matched filtering can be used to make quantitative
tests of general relativity in a highly dynamical and nonlinear
regime, and to make measurements of the binary's parameters.  These
measurements and tests can be carried out with moderate accuracy by
LIGO/VIRGO, and with extremely high accuracy by the proposed
space-based interferometer LISA.  Using information theory, we
estimate the total number of bits of information obtainable from the
merger waves ($\sim 10$ to $60$ bits for LIGO/VIRGO, up to $\sim 200$
bits for LISA), and estimate how much information would be lost due to
numerical errors in the templates or to sparseness in the template
grid.  We deduce an approximate rule-of-thumb for the required
accuracy of merger templates and for their spacing.

\end{abstract}
}

\def\mvec{\vec}
\def\bftheta{\mbox{\boldmath $\theta$}}

\narrowtext

\section{INTRODUCTION AND SUMMARY}
\label{intro}

\subsection{Gravitational waves from binary black hole systems}
\label{introsuboverview}

With the kilometer-scale, ground-based interferometric
gravitational-wave observatories LIGO~\cite{ligoscience},
VIRGO~\cite{virgo}, and GEO600 \cite{geo} expected to be on line and
taking data within the next few years, and with the space-based
interferometer LISA~\cite{cornerstone,lisa,LISAreport} in the planning
and development stage, much effort is currently going into
understanding potential gravitational-wave sources and associated data
analysis issues.  One potentially very interesting and important class
of source is the coalescence of binary black holes (BBHs) where the
two black holes have comparable masses.  Such binaries with total
masses $M$ in the range $10 M_\odot \alt M \alt 10^3 M_\odot$ could be
detected by ground-based interferometers, and with $10^5 M_\odot \alt
M \alt 10^8 M_\odot$ by LISA. 

The evolution of these systems, and the gravitational waves that they
emit, can be roughly divided into three successive epochs: an
adiabatic {\it inspiral} epoch, in which the evolution of BBH systems
is driven by radiation reaction, and which terminates roughly at the last
stable circular orbit\ \cite{kww,cook}; a violent, dynamical {\it
merger} epoch; and a {\it
ringdown} epoch in which the emitted gravitational waves are dominated
by the $l=m=2$ quasinormal mode radiation of the final Kerr black
hole.  Gravitational waves from the merger phase could be rich with
information about relativistic gravity in a highly nonlinear, highly
dynamical regime which is poorly understood today.

Theoretical predictions of the gravitational waveforms $h_+(t)$ and
$h_\times(t)$ produced in the three phases of BBH coalescences will be
useful both for detecting the gravitational-wave signal, and for interpreting
and making deductions from the observed waveforms, {\it i.e.}, for
extracting information from the waves. 

For the inspiral phase, such theoretical waveforms or waveform
templates have already been computed analytically to
post-2.5-Newtonian order \cite{Blanchet,ordercount}.  These templates
will be accurate enough for separations $r\agt 12M$ that their errors
will not significantly impede wave detection; for more details see,
for example, Ref.\ \cite{paperI} and Sec.\ \ref{priorinsp} below.  The
phase evolution of the inspiral waves between $r \sim 12 M$ and $r \sim 6
M$, where $M$ is the total mass of system and $r$ the distance between
the black holes in Schwarzschild coordinates, will not be accurately
described by the post-Newtonian approximation \cite{PNbreakdown}.
Alternative analytic and numerical approximation schemes are under
development for modeling the coalescence and computing the waves in
this ``Intermediate binary black hole'' (IBBH) regime
\cite{Price,Pat,Detweiler}.  For the purpose of this paper, we
consider this IBBH regime to be part of the inspiral phase of the
coalescence.

Templates for the ringdown phase of the coalescence are obtained using
perturbation theory on the background of the final Kerr black hole
\cite{chandra}; these templates consist of exponentially damped
sinusoids.

In contrast to the situation for the inspiral and ringdown phases,
there is at the present time very little theoretical understanding of
gravitational waves from the merger phase, and no merger templates
exist at all.  Detailed understanding of the merger probably will come
only from numerical relativity.  One rather large effort to compute
the dynamics of BBH mergers is the American Grand Challenge Alliance,
an NSF funded collaboration of physicists and computer scientists at
eight institutions \cite{grandchall,samreview}; similar efforts are
underway elsewhere.  Modeling BBH mergers is an extremely difficult
task; the numerical relativists who are writing codes for simulating
BBH mergers are beset with many technical difficulties.

Our theoretical understanding of BBH mergers could be in any one of
several different states by the time the first BBH coalescences are
detected: (i) {\it No information:} The supercomputer simulation codes
have not yet been successfully implemented, thus no information about
waves from BBH mergers is available.  (ii) {\it Information limited in
principle:} A small amount of information about the waves is
available.  This could arise if working supercomputer codes are
available, but the codes cannot simulate fully general BBH mergers,
but only those in some special class ({\it e.g.}, vanishing initial
spins, or equal mass black holes).  Or, it could arise if the codes
can simulate arbitrary mergers but technical difficulties prevent the
extraction of accurate gravitational waveforms; in such cases one
would know at least the duration of the merger waves.  (iii) {\it
Information limited in practice:} Fully general BBH mergers can be
simulated and waveforms can be extracted, but each run of these codes
to produce a template is very expensive in terms of computer time and
cost, and therefore only a small number of representative template
shapes can be computed and stored.  (The total number of template
shapes required to cover the entire range of behaviors of BBH mergers
is likely to be in the range of thousands to millions or more.)  (iv)
{\it Full information:} A complete set of theoretical templates has
been computed and is available for data analysis.  This fourth
possibility seems rather unlikely in the time frame of the first
detections of BBH coalescences.

\subsection{Detecting the waves}
\label{introsubdetection}

Depending on the system's mass, some BBH coalescence events will be
most easily detected by searching for the inspiral waves, others by
searching for the ringdown waves, and others by searching for the
merger waves themselves (depending on the systems mass).  In paper I
of this series \cite{paperI}, we analyzed the prospects for detecting
BBH events using these three different types of searches, for initial
and advanced LIGO interferometers and for LISA.  We briefly review
here some of the relevant aspects and conclusions of that analysis.

Low-mass BBHs [$M \alt 30 M_\odot$ for the first LIGO interferometers;
$(1+z) M \alt 80 M_\odot$ for the advanced LIGO interferometers;
$(1+z) M \alt 3 \times 10^6 M_\odot$ for LISA, where $z$ is the
source's cosmological redshift] are best searched for via their
inspiral waves.  Such searches will use matched filtering with
post-Newtonian templates.  These low-mass binaries may be the most
common type of detected BBH source.  Moreover, they may well be the
first detected source of gravitational waves and be detected before
binary neutron star inspirals, since the range of initial LIGO
interferometers for BBHs with $M \alt 50 M_\odot$ is $\sim 250 \, {\rm
Mpc}$ whereas binary neutron stars can be seen out to $\sim 25 \, {\rm
Mpc}$ \cite{paperI,rangenote}.

Higher mass BBH systems are best searched for via their ringdown waves
or merger waves.  A matched filtering search for ringdown waves will
be possible as soon as data are available, since ringdown templates are
simple to construct.

A matched filtering search for merger waves could be performed {\it
if} a complete set of merger templates were available.  We estimated
in Ref.~\cite{paperI} that the resulting event detection rate would be
a factor of roughly $40$ higher than the event rate from inspiral and
ringdown searches for a certain range of BBH masses ($30 M_\odot \alt
M \alt 200 M_\odot$ for initial LIGO interferometers, $100 M_\odot
\alt M \alt 400 M_\odot$ for advanced LIGO interferometers, and $3
\times 10^6 M_\odot \alt (1+z) M \alt 3 \times 10^7 M_\odot$ for
LISA).  However, as mentioned above in Sec.\ \ref{introsuboverview},
it seems very unlikely that a complete bank of numerical templates
will be available.  If merger templates are not available, one can
still search for the merger waves using simple band-pass filtering
({\it i.e.}, using filters that throw away all signal and noise except for
that within some prescribed frequency band), or more effectively using
techniques such as the noise-monitoring search method described in
Refs.\ \cite{paperI,ComingSoonToAJournalNearYou}.  The gain factor in
event detection rate for noise-monitoring searches for merger waves,
over inspiral and ringdown searches, will be roughly $4$ to $10$,
depending on (among other things) whether or not one has firm
information from representative supercomputer simulations about the
possible durations and frequency bandwidths of merger waveforms
\cite{noterate}.

Once a BBH event has been detected, the location of the three
different phases of the waves in the data stream will be known to a
fair approximation.  For many detected events, though, it will not be
the case that all three phases will be detectable.  For instance,
typical low mass BBH events which are detected via their inspiral
waves will have ringdown waves that are too weak to be detected; see
Ref.\ \cite{paperI} and Sec.\ \ref{priorinsp}.  Likewise, very massive
systems which are detected via their ringdown waves might in some
cases not yield a detectable inspiral signal.

\subsection{Extracting the waves' information: three scenarios}
\label{extractionsec}

In contrast to paper I \cite{paperI}, where we focused on expected
signal strengths and search strategies for BBH events, in this paper
we focus on measurements of the merger waveform itself: on
reconstructing the waveform from the instrumental data stream, and on
using the measured waveforms to learn about the BBH source and about
the dynamics of very strong field general relativity.  At present,
because merger waveforms are so poorly understood, it is hard to say
how much one can learn about BBH systems from their merger waves.
Both how well we can reconstruct BBH waveforms and how much we can
learn from such reconstructions depend on the success of efforts to
numerically simulate BBH mergers.

In this subsection, as background to the discussion of the contents of
this paper in Sec.~\ref{overviewpaper} below, we describe in general
terms three possible different scenarios for data analysis of the
merger waves:

The first possibility [corresponding to situation (i) in Sec.\
\ref{introsuboverview}] is that numerical computations might provide no
input at all that can be used to aid gravitational-wave data analysis.
In this case, with no templates to guide the interpretation of the
measured waveform, it will not be possible to obtain any information
about the BBH source or about strong-field general relativity from the
merger waves.  One's goal will simply be to measure as accurately as
possible the merger waveform's shape.  For this waveform shape
measurement, observers should make use of all possible prior
information obtainable from analyses of the inspiral and/or ringdown
signals, if they are detectable.  (For example, if the system is
detected via its inspiral waves, then one will know that the merger
waves lie immediately following the inspiral waves in the data stream,
and must join smoothly onto the ringdown waves.)

Second [situations (ii) and (iii) of Sec.\ \ref{introsuboverview}], if
one has only a few, representative supercomputer simulations and
associated waveform templates at one's disposal, one might simply
perform a qualitative comparison between the measured waveform and
templates in order to deduce qualitative information about the BBH
source.  For instance, simulations might demonstrate a strong
correlation between the duration of the merger (in units of the total
mass of the system) and the spins of the black holes in the binary.
One might then be able to deduce some information about the black hole
spins from the duration of the reconstructed merger waveform, without
having to find a template that exactly matched the measured waveform.
In this second scenario, for the purpose of reconstructing a ``best
fit'' merger waveform from the noisy data stream, one should use the
prior information from the measured inspiral and/or ringdown waves,
and in addition the prior information (for example the expected range
of frequencies) one has about the merger waveforms' behaviors from the
representative supercomputer simulations.

The third scenario consists of performing matched filtering analyses
of the data stream with merger templates in order to measure the
parameters of the BBH binary and to test general relativity.  This
will certainly be feasible if one has a complete set of merger
templates [situation (iv) of Sec. \ref{introsuboverview}].  However,
in some cases matched filtering parameter extraction may also be
feasible in situation (iii) of Sec.~\ref{introsuboverview},
where one has a working computer code for simulating BBH mergers but
where each run of the code is so expensive in computer time and cost
that it is not possible to calculate a complete set of templates.  In
such a case, after the merger waves have been detected, it may be
possible to perform several runs of the supercomputer code,
concentrated in the appropriate small region of parameter space
compatible with one's measurements from the inspiral and ringdown
waves, in an effort to match the observed waveforms.  In either case
(complete set of templates or templates produced as needed), a
conclusive fit between a numerical waveform and the measured waveform
would be a triumph for general relativity, testing the theory in an
extremely strong field, fast motion regime with no approximations, and
would provide an unequivocal signature of the existence of black
holes.

In this paper, as we now outline, we consider the requirements for and
the implications of all three of these modes of data analysis. 

\subsection{Extracting the waves' information: our analyses, suggested
tools, and results}
\label{overviewpaper}

The four principal purposes of this paper are: (i) to review and
discuss the useful information carried by all three phases of the
waves and the prospects for its extraction, both with and without
templates; (ii) to suggest a data analysis method that can be used in
the absence of templates to obtain from the noisy data stream a
``best-fit'' merger waveform shape; (iii) to provide input to
numerical relativity simulations by highlighting the kinds of
information that supercomputer simulations can provide, other than
merger templates, that can aid BBH merger data analysis; and (iv) to
provide input to numerical relativity simulations by deriving some
requirements that numerical templates must satisfy in order to be as
useful as possible for data analysis purposes.  We now turn to a
detailed summary of our analyses and results in these four areas.

We first consider the situation in which very little information about
the merger waveform is available to aid data analysis.  The data
analysis method that we suggest [item (ii) in the above paragraph]
reduces in this case to band-pass filtering.  In this case, observers
will likely resort to simple band-pass filters to study the merger
waves.  The first question to address in this context is whether the
merger signal is likely to even be {\it visible}; that is, whether the
signal will stand out above the background noise level in the
band-pass filtered detector output.

The merger signal will be visible if the band-pass filtering
signal-to-noise ratio (SNR) is large compared to unity.  In paper I of
this series, we estimated the matched filtering SNRs that could be
obtained from the merger signal if templates were available ({\it cf.}
Figs.\ 4, 5, and 6 of Ref.~\cite{paperI}); and we estimated that the
SNRs that can be achieved for the merger signal with band-pass filters
will be roughly a factor of 5 smaller than the matched filtering SNRs.
The resulting values of band-pass filtering SNR depend on the distance
to the BBH.  In Sec.~\ref{visibility} we estimate the distance to
typical BBHs with $M \alt 20 M_\odot$ that have been detected via
their inspiral signals by initial LIGO interferometers, and we infer
that the merger signal is likely to be marginally visible (band-pass
filtering SNR $\sim 1$) for typical detected events.  For advanced
LIGO interferometers, we estimate that the merger signal is somewhat
less likely to be visible (band-pass filtering SNR $\sim 1/4$).  The
reason for this somewhat counterintuitive result is that matched
filtering is more efficient, relative to band-pass filtering, for
advanced interferometers.  Thus, only the somewhat rarer, stronger
merger signals will be visible for advanced LIGO interferometers.  For
LISA, by contrast, we estimate that the band-pass filtering SNRs will
typically be $\agt 200$ and thus the merger waves will easily be
visible.

In Sec.~\ref{inspiralvisibility}, for comparison, we estimate the
band-pass filtering SNRs of the last few cycles of inspiral waves
({\it i.e.}, just before merger) and find them to be typically of
order unity for low mass BBH events detected by ground-based
interferometers.  Thus, the last few cycles of the inpiral should be
(just about) individually visible above the interferometer noise.

When templates are not available, one's goal will be to reconstruct as
well as possible the merger waveform from the noisy data stream.  In
Sec.\ \ref{bestguess} we use Bayesian statistics and the framework of
maximum likelihood estimation to sketch out an optimized method for
performing such a reconstruction in the absence of theoretical
templates.  The method is based on a ``perpendicular projection'' of
the observed noisy signal onto an appropriate function space that
encodes all our (possibly sketchy) prior knowledge about the
waveforms.  We argue that the best type of ``basis functions'' to use
to specify this function space are wavelets: functions which
simultaneously allow localization in time and in frequency.  We
develop this reconstruction technique in detail using the language of
wavelets.  We show that the operation of ``perpendicular projection''
into the function space is a special case of Wiener optimal filtering.
In Sec.~\ref{fidelity} and Appendix \ref{correl} we demonstrate
mathematically the rather obvious result that the reconstructed signal
will statistically be a good representation of the true signal (as
measured by a correlation integral between the true signal and
reconstructed signal) only in the regime where the band-pass filtering
SNR is large.

In Appendix \ref{network}, we describe an extension of the method to a
network of several gravitational wave detectors which allows one to
reconstruct, from the outputs of all the detectors in the network, the
two independent waveforms $h_+(t)$ and $h_\times(t)$ of the merger
waves.  We also show that our method for a network is an extension and
generalization of a method previously suggested by G\"ursel and Tinto
\cite{tinto}.  Secs.~\ref{deriveposterior} and \ref{direction} of Appendix
\ref{network} overlap somewhat with unpublished analyses by Sam Finn
\cite{FinnRecent}.  Finn uses similar mathematical techniques to analyze
the use of multiple interferometers to measure a stochastic background
of gravitational waves and to measure waves of well-understood form,
applications which are rather different from the measurement of bursts
of unknown form that we consider.

Our waveform reconstruction algorithm comes in two versions:  a simple
version incorporating the above mentioned ``perpendicular
projection'', described in Sec.~\ref{deriveBG}, and a more general and
powerful version that allows one to build in more prior information,
described in Sec.~\ref{extendedmethod}.  If one's prior information
consists only of knowledge about the signal's bandwidth, then the
best-fit reconstructed waveform is just the band-pass filtered
data stream.  However, one can also build in as input to the method
the expected duration of the signal, the fact that it must match up
smoothly to the measured inspiral waveform, {\it etc.}; in such cases
the reconstructed waveform differs from the band-pass filtered data
stream.  

Qualitative information about BBH merger waveforms will thus be very
useful as prior information for signal reconstruction.  Such
information will also be useful as a basis for qualitative comparisons
with the reconstructed waveforms in order to make qualitative
deductions about the BBH source, as outlined in
Sec.~\ref{extractionsec} above.  Supercomputer simulations should be
able to provide such information, in the case where these codes can
successfully simulate BBH mergers and produce templates, but where
running the codes is too expensive to permit an extensive survey of
the merger parameter space ({\it i.e.}, too expensive to produce a
complete set of templates).  In this situation, a small number of
representative simulations could still be extremely useful.  In Sec.\
\ref{notmplsuper}, we give examples of the types of information such
supercomputer simulations could provide (short of providing a complete
set of merger templates): the range of numbers of cycles in the merger
waveform and how this number depends on parameters such as the initial
spins of the black holes; the (closely related) range of temporal
durations of merger waveforms, and how duration varies with parameters
of the binary; the minimum and maximum frequencies of typical merger
energy spectra; characteristics of the waveform's time/frequency
behavior (whether it involves a monotonic chirp or not, and whether in
some cases it can be characterized as a modulated carrier wave or
not); and which quasinormal modes are typically excited, and how
strongly.

We turn next to issues concerning the use of numerical templates in
data analysis.  In Sec.\ \ref{infowithtemplates}, we begin to examine
matched filtering of merger waves with templates.  As mentioned in
Sec.~\ref{extractionsec} above, such matched filtering may be possible
even if a complete set of merger templates does not exist: runs of
merger template generation codes can be performed as part of the data
analysis of measured BBH signals in an effort to produce a template
that matches the measured waveform; such efforts may or may not be
successful.  We review in Sec.~\ref{infowithtemplates} what one should
be able to achieve with matched filtering: measurements of the
binary's physical parameters (masses, vectorial spin angular momenta,
{\it etc.}) which are independent of any such measurements from the inspiral
and ringdown waves; and quantitative tests of general relativity in
the most extreme of domains: highly nonlinear, rapidly dynamical,
highly non-spherical spacetime warpage.  These measurements and tests
will be possible with modest accuracy with LIGO/VIRGO, and with
extremely high accuracy with LISA (for which the merger matched
filtering SNRs are typically $\agt 10^4$ \cite{paperI}).

In order for such measurements and tests to be as successful as
possible, the numerically generated merger templates must satisfy
certain requirements.  In Sec.~\ref{tmplacc} we derive a simple
formula [Eq.~(\ref{accuracycriterion})] that numerical relativists can
use to ensure that the waveforms produced by their simulations are
sufficiently accurate for data analysis.  In Sec.~\ref{accimpl} we
describe how this formula can be used to regulate the accuracy with
which the numerical simulations are carried out.  The formula is
derived from the following requirements: first, any signal searches
that use matched filtering with merger templates should suffer a
fractional loss of event rate due to template inaccuracies of no more
than $3 \%$; and second, when using templates to fit for and measure
the physical parameters of the BBH source (masses, spins {\it etc.}),
the systematic errors due to template inaccuracies should always be
smaller than the detector-noise induced statistical errors.  The
derivation of the formula from these two requirements is given in
Sec.~\ref{derivationacc}.

In Sec.~\ref{infosec}, we address again the issue of template accuracy
requirements, and also the issue of the required spacing of templates
in parameter space in the construction of a grid of templates, by
using the mathematical machinery of information theory.  In
information theory, a quantity called ``information'' (analogous to
entropy) can be associated with any measurement process: it is simply
the base 2 logarithm of the number of distinguishable outcomes of the
measurement \cite{brillouin,coverthomas}.  Equivalently, it is the
number of bits required to store the knowledge gained from the
measurement.  We specialize the notions of information theory to
gravitational wave measurements, and define two different types of
information: (i) the ``total'' information $I_{\rm total}$ which is
the base 2 logarithm of the total number of distinguishable waveform
shapes that the measurement could have produced; and (ii) a smaller
``source'' information $I_{\rm source}$, which is the base 2 logarithm
of the total number of distinguishable waveform shapes that the
measurement could have produced {\it and} that are generated by BBH
mergers.  This second measure of information is equivalent to the base
2 logarithm of the total number of independent BBH sources that the
measurement could have distinguished.  We give precise definitions of
these two notions of information [Eqs.~(\ref{infodef}) and
(\ref{infodef1})] in Sec.\ \ref{infosec}.  In Appendix \ref{app_info},
we derive simple analytic approximations for the quantities $I_{\rm
total}$ and $I_{\rm source}$ [Eqs.~(\ref{infoapprox}) and
(\ref{infoapprox1})], expressing them in terms of the merger signal's
matched filtering signal-to-noise ratio $\rho$, the number of
independent real data points ${\cal N}_{\rm bins}$ in the observed
signal, and the number of parameters ${\cal N}_{\rm param}$ on which
merger templates have a significant dependence.  We estimate that the
total information gain $I_{\rm total}$ is typically of the order of
$\sim 10$ to $\sim 120$ bits for LIGO/VIRGO, and can be up to $\sim
400$ bits for LISA; and that the source information gain $I_{\rm
source}$ is typically of the order of $10$ to $70$ bits for
LIGO/VIRGO, and can be up to $\sim 200$ bits for LISA.

In Sec.~\ref{infolosssec}, we estimate the loss in information about
the BBH source, $\delta I_{\rm source}$, that would result from
template inaccuracies [Eq.~(\ref{anss}) below]; this allows us to
re-derive the criterion for the template accuracy requirements
obtained in Sec.~\ref{tmplacc}.  We also estimate the loss in
information $\delta I_{\rm source}$ that would result from having
insufficiently closely spaced templates in a template grid
[Eq.~(\ref{anss3}) below], and we deduce an approximate criterion for
how closely templates must be spaced.

\subsection{Organization of this paper}
\label{introsuborg}

The remainder of this paper is organized as follows: In Sec.\
\ref{notations}, we define the notations and conventions that we will
use throughout the paper.  In Sec.\ \ref{priorinsp}, we review in
moderate detail the information obtainable from the inspiral and
ringdown phases of the waves for detected BBH events, which will be
used as prior information when attempting to analyze the merger phase.

In Sec.\ \ref{visibility}, we discuss the visibility of BBH
coalescence waveforms.  In Sec.\ \ref{inspiralvisibility}, we first
compute the band-pass filtering SNR for the last few cycles of the
inspiral; this serves as background to the merger visibility analysis,
and is relevant to the merger visibility itself: if the end of
the inspiral is visible, then the beginning of the merger will most likely
be visible as well.  In Sec.\ \ref{mergervisibility}, we analyze the
merger visibility.

In Sec.\ \ref{bestguess} we present our method for optimally
reconstructing the merger waveform from the interferometer output.  We
derive the method in Sec.\ \ref{deriveBG}, and in Appendix
\ref{network} we present an extension of the method to a network of
several gravitational-wave detectors.  In Sec.\
\ref{extendedmethod} we describe another extension of the method 
that allows one to incorporate prior information in a more effective
way.  In Sec.\ \ref{fidelity} we quantify the fidelity of the
reconstructed waveform by defining a normalized correlation
coefficient that describes how well the reconstructed wave correlates
with the true waveform.  We show in Appendix \ref{correl} that this
coefficient will be close to 1 ({\it i.e.}, that the reconstructed
waveform will be close to the true waveform) when the signal's
band-pass filtering SNR is $\gg 1$.

In the remainder of the paper, we consider the situation where
supercomputer simulations are able to provide some input to data
analysis, either in the form of useful qualitative or
semi-quantitative information about the merger, or in the form of
templates.  Sec.\ \ref{notmplsuper} presents a list of the kinds of
information that numerical relativists may be able to provide, short
of a definitive template set, that can be used to aid data analysis.
Sec.\ \ref{infowithtemplates} discusses and describes the kinds of
information that can be obtained from the gravitational wave data when
merger templates are available.

In Secs.\ \ref{tmplacc} and \ref{infosec} we present our derivations
of criteria for determining whether templates are numerically accurate
enough and closely spaced enough to be used in data analysis.  In
Sec.\ \ref{derivationacc} we derive an accuracy criterion from the
requirement that the loss in event detection rate due to template
inaccuracies in a matched filtering signal search using merger
templates be no more than $3 \%$.  We also obtain, in
Sec.~\ref{derivationacc}, approximately the same criterion from
demanding that systematic errors in parameter extraction using merger
waveforms be small compared to the detector-noise induced statistical
errors.  In Sec.\ \ref{infolosssec} we rederive the accuracy criterion
using the mathematical machinery of information theory.  In this
derivation, we require that the number of bits of information lost due
to template inaccuracies be less than $1$.  The relevant information
theoretic concepts are presented in Secs.~\ref{Itotalsec} and
\ref{Isourcesec}; some of the technical calculations are relegated to
Appendix \ref{app_info}.

Finally, in Sec.\ \ref{concl} we summarize our main conclusions.

\section{Notations and conventions}
\label{notations}

In this section we introduce some of the conventions and notations that
will be used throughout the paper.  We use geometrized units in which
Newton's gravitational constant $G$ and the speed of light $c$ are
unity.  For any function of time $a(t)$, we will use a tilde to represent
that function's Fourier transform, according to the convention
\begin{equation}
{\tilde a}(f) = \int_{-\infty}^\infty dt\,e^{2\pi i f t} a(t).
\label{fourierdef}
\end{equation}
The output strain amplitude $s(t)$ of a gravitational wave detector
can be written as
\begin{equation}
s(t) = h(t) + n(t),
\label{decompos1}
\end{equation}
where $h(t)$ is the gravitational wave signal and $n(t)$ is the
detector noise.  Throughout this paper we will assume, for simplicity,
that the noise is stationary and Gaussian.  The statistical properties
of the noise determine a natural inner product $\left( \ldots | \ldots
\right)$ on the vector space of waveforms $h(t)$, given by
\begin{equation}
\left( h_1 \,| \, h_2 \right) = 4 \, {\rm Re} \, \int_0^\infty df
{{{\tilde h}_1(f)^* \, {\tilde h}_2(f)}\over S_h(f)};
\label{innerprod}
\end{equation}
see, for example, Refs.~\cite{finn2,cutlerflan94}.  In
Eq.~(\ref{innerprod}), $S_h(f)$ is the power spectral density of the
strain noise $n(t)$ \cite{twosided}.
The associated norm is given by
\begin{equation}
||h|| \equiv \sqrt{ \left( h \, | \, h \right)}.
\label{norm}
\end{equation}
For any waveform $h(t)$, the matched filtering signal-to-noise ratio
is given by 
\begin{equation}
\rho^2 = \left( h \, | \, h \right)  = 4 \int_0^\infty df \, {|
{\tilde h}(f)|^2 \over S_h(f)}.
\label{snr}
\end{equation}

On several occasions we shall be interested in finite stretches of
data of length $T$ say, represented in a discrete way as a
vector of numbers instead of as a continuous function.  If $\Delta t$
is the sampling time, this vector is
\begin{equation}
{\bf s} = (s^1, \ldots, s^{{\cal N}_{\rm bins}})
\label{bfnotation}
\end{equation}
where ${\cal N}_{\rm bins} = T/\Delta t$, $s^j = s(t_{\rm
start} + j \Delta t)$, $0 \le j \le {\cal N}_{\rm bins}$, and $t_{\rm
start}$ is the starting time.
The quantity ${\cal N}_{\rm bins}$ is
the number of independent real data points (number of bins) in the
measured signal; it is denoted by ${\cal N}$ in Appendices
\ref{network} and \ref{correl}.  The gravitational wave signal $h(t)$
and the noise $n(t)$ can similarly be 
represented in this way, so that ${\bf s} = {\bf h} + {\bf n}$, as in
Eq.~(\ref{decompos1}).  We adopt the geometrical viewpoint of
Dhurandhar and Schutz \cite{DS}, regarding ${\bf s}$ as an element of
an abstract vector space $V$ of dimension ${\cal N}_{\rm bins}$, and
the sample points $s^j$ as the components of ${\bf s}$ on a time
domain basis $\{{\bf e}_1, \ldots, {\bf e}_{{\cal N}_{\rm bins}}\}$ of
$V$:
\begin{equation}
{\bf s} = \sum_{j=1}^{{\cal N}_{\rm bins}}\, s^j \, {\bf e}_j.
\end{equation}
Taking a finite Fourier transform of the data stream can be regarded
as a change of basis of $V$ in which ${\bf s}$ remains fixed but its
components change.  Thus, a frequency domain basis $\{{\bf d}_k\}$ of
$V$ is given by the finite Fourier transform
\begin{equation}
{\bf d}_k = \sum_{j=1}^{{\cal N}_{\rm bins}} \, {\bf e}_j \exp \, \{2 \pi i j
k/{{\cal N}_{\rm bins}}\},
\label{freqbasis}
\end{equation}
where $-({\cal N}_{\rm bins}-1)/2 \le k \le ({\cal N}_{\rm
bins}-1)/2$.  The corresponding 
frequencies $f_k = k/T$ run from $-1/(2 \Delta t)$ to $1/(2
\Delta t)$ {\cite{note4}}.

More generally, if we band-pass filter the data stream down to a
frequency interval of length $\Delta f$, and consider a stretch of
band-pass filtered data of duration $T$, this stretch of data will
have  
\begin{equation}
{\cal N}_{\rm bins} = 2 T \Delta f
\end{equation}
independent real data points.  In this case also we regard the set of
all such stretches of data as an abstract linear space $V$ of
dimension ${\cal N}_{\rm bins}$.

On an arbitrary basis of $V$, we define the matrices $\Gamma_{ij}$ and
$\Sigma^{ij}$ by
\begin{equation}
\langle n^i \, n^j \rangle = \Sigma^{ij}
\label{Sigmadef}
\end{equation}
and 
\begin{equation}
\Gamma_{ij} \, \Sigma^{jk} = \delta^i_k;
\label{Gammadef}
\end{equation}
{\it i.e.}, the matrices ${\bf \Gamma}$ and ${\bf \Sigma}$ are
inverses of each other.  In Eq.~(\ref{Sigmadef}) the angle brackets
mean expected value.  On the time domain basis $\{ {\bf e}_1, \ldots,
{\bf e}_{{\cal N}_{\rm bins}} \}$, we have
\begin{equation}
\Sigma^{jk} = C_n( t_j - t_k),
\label{SigmadefT}
\end{equation}
where $t_j = t_{\rm start} + j \Delta t$, and $C_n(\tau)  
= \langle n(t) n(t+ \tau) \rangle$ is the noise correlation function
given by 
\begin{equation}
\label{c_n_def}
C_n(\tau) = \int_0^{\infty} d f \, \cos[2 \pi f \tau] \, S_h(f).
\end{equation}
We define an inner product on the space $V$ by
\begin{equation}
\left({\bf h}_1 \,|\,{\bf h}_2\right) = \Gamma_{ij} h_1^i h_2^j
\label{discreteinnerprod}
\end{equation}
This is essentially a discrete version of the inner product
({\ref{innerprod}}) which characterizes the detector noise: the two
inner products coincide in the limit of small sampling times $\Delta
t$, and for waveforms which vanish outside of the time interval of
length $T$
\cite{finnmeasure}.

Throughout this paper we shall use interchangeably the notations
$h(t)$ and ${\bf h}$ for a gravitational waveform.  We shall also for
the most part not need to distinguish between the inner products
(\ref{innerprod}) and (\ref{discreteinnerprod}).  Some generalizations
of these notations and definitions to a network of several detectors
are used in Appendix \ref{network}.

For a given detector output ${\bf s} = {\bf h} + {\bf n}$, we define
\begin{equation}
\rho({\bf s})^2 = || {\bf s}||^2 = \left( {\bf s} \, | \, {\bf s}
\right),
\label{magnitudedef}
\end{equation}
which is the inner product or integral of the detector output with
itself.  We will call $\rho({\bf s})$ the {\it magnitude} of the
stretch of data ${\bf s}$.
From Eqs.~(\ref{Sigmadef}) and (\ref{discreteinnerprod}) it
follows that 
\begin{equation}
\langle \rho({\bf s})^2 \rangle = \rho^2 + {\cal N}_{\rm bins},
\label{magnitude1}
\end{equation}
where $\rho^2$ is the matched filtering SNR squared (\ref{snr}) of the
signal ${\bf h}$, and 
that
\begin{equation}
\sqrt{\left< \, [\Delta \rho({\bf s})^2]^2 \,\right>} = \sqrt{4 \rho^2
+ 2 {\cal N}_{\rm bins}},
\label{magnitude2}
\end{equation}
where $\Delta \rho({\bf s})^2 \equiv \rho({\bf s})^2 - \langle
\rho({\bf s})^2 \rangle$.
Thus, the magnitude $\rho({\bf s})$ is approximately the same as the
usual SNR $\rho$ in the limit $\rho \gg \sqrt{{\cal N}_{\rm bins}}$
(large signal-to-noise squared per frequency bin), but is much larger
than $\rho$ when $\rho \ll \sqrt{{\cal N}_{\rm bins}}$.  The quantity
$\rho({\bf s})$ will occur in our in our information theory
calculations in Sec.~\ref{infosec} and Appendix \ref{app_info}.

The space $V$ equipped with the inner product
(\ref{discreteinnerprod}) forms a Euclidean vector space.  We will
also be concerned with sets of gravitational waveforms ${\bf
h}({\bftheta})$ [equivalently, $h(t; {\bftheta})$] that depend on a
finite number $n_p$ of parameters ${\bftheta} = (\theta^1, \ldots,
\theta^{n_p})$.  For example, inspiral gravitational waveforms form a
set of this type, where ${\bftheta}$ are the parameters describing the
binary source.  We will denote by ${\cal S}$ the manifold of signals
${\bf h}({\bftheta})$, which is a submanifold of dimension $n_p$ of
the vector space $V$.  We will adopt the convention that Roman indices
$i$, $j$, $k$, \ldots will run from 1 to ${\cal N}_{\rm bins}$, and
that $v^i$ will denote some vector in the space $V$.  Greek indices
$\alpha$, $\beta$, $\gamma$ will run from $1$ to $n_p$, and a vector
$v^\alpha$ will denote a vector field on the manifold ${\cal S}$.  The
inner product (\ref{discreteinnerprod}) induces a natural Riemannian
metric on the manifold ${\cal S}$ given by
\begin{equation}
ds^2 = \left( {\partial {\bf h} \over \partial \theta^\alpha} \,
\bigg| \, {\partial {\bf h} \over \partial \theta^\beta }\right) \,\,
d\theta^\alpha d\theta^\beta.
\label{metricTH}
\end{equation}
We shall denote this metric by $\Gamma_{\alpha\beta}$ and its inverse
by $\Sigma^{\alpha\beta}$, relying on the index alphabet to
distinguish these quantities from the quantities (\ref{Sigmadef}) and
(\ref{Gammadef}).  For more details on this geometric picture, see,
for example, Ref.~\cite{cutlerflan94}.

We shall use the word {\it detector} to refer to either a single
interferometer or a resonant mass antenna, and the phrase {\it
detector network} to refer to a collection of detectors operated in
tandem.  Note that this terminology differs from that adopted in, for
example, Ref.~\cite{finn2}, where a detector network is called simply
a detector.

Finally, we will use bold faced vectors like ${\bf a}$ to denote
either vectors in three dimensional space, or vectors in the ${\cal
N}_{\rm bins}$-dimensional space $V$.  In Appendix \ref{network}, we
will use arrowed vectors ({\it e.g.}, ${\vec a}$) to denote elements
of the linear space of the output of a network of gravitational wave
detectors.

\section{INFORMATION FROM THE INSPIRAL AND RINGDOWN PHASES}
\label{priorinsp}

Different types of information will be obtainable from the three
different phases of the gravitational wave signal.  If the inspiral
and ringdown phases are strong enough to be measurable, they will be
easier to analyze than the merger phase, and the information they
yield will be used as ``prior information'' in attempting to analyze
the merger phase.  For instance, from the inspiral portion of the
signal it will be possible to measure the masses of the binary's black
holes to some accuracy (as we discuss below).  Those measured masses
will then be an input to data analysis of the merger waves, since they
strongly constrain the possible values of template parameters
that need to be examined when fitting a theoretical waveform to the
merger signal.  In this section, we review the prior information that
will likely be available from measurements of the inspiral and the
ringdown in typical cases.

Let us focus first on solar mass coalescences [$(1+z) M \alt 50
M_\odot$ say] measured by ground based interferometers, for which most
of the prior information will come from the inspiral waveforms.  The
analysis of the inspiral waveforms will take place in two phases.  The
first phase will consist of filtering the data streams of each detector
separately using ``search templates'' in order to detect the
inspiral~\cite{notedetect}.  These search templates will depend on $2$
or possibly $3$ parameters.  Roughly $10^4$ to $10^5$ distinct
template shapes will be required for initial LIGO interferometers, and
roughly $10^6$ to $10^7$ template shapes for advanced LIGO
interferometers
\cite{owen,prl,haristmpl,searchtemplates,krolaknew,insptmplcount}.
(Note that these numbers assume that the search is for generic
inspiraling binaries, not simply black hole binaries.  If the search
were restricted to BBH systems only, these numbers would be greatly
reduced: assuming that the smallest BBH systems consist of a pair of
$2 M_\odot$ binaries, the number of templates for initial LIGO
interferometers is roughly $10^3$, and for advanced interferometers
roughly $10^5$.)  The second phase will consist of combining the
outputs of all the detectors together and using the most accurate
templates available (``extraction templates'') to analyze the signal
and extract the best-fit parameter values.  Such extraction templates
will presumably be provided by post-Newtonian calculations, perhaps
improved by the judicious use of Pad\'e approximants \cite{Pade}, and
perhaps supplemented by IBBH calculations in the IBBH regime $6 M \alt
r \alt 12 M$ (cf.\ the discussion in Sec.~\ref{introsuboverview}
above).  In this second phase there will be 15 independent parameters
to fit for.  These parameters are the masses $m_1$ and $m_2$ and
initial spins ${\bf S}_1$ and ${\bf S_2}$ of the two black holes, the
luminosity distance $D$ to the binary, the direction of the orbital
angular momentum ${\hat {\bf L}} = {\bf L} / | {\bf L} |$, the
direction ${\hat {\bf n}}$ from the binary to the Earth, and the
arrival time $t_c$ and orbital phase $\phi_c$ at some fiducial
frequency.  (The dependence of the templates on several of these 15
parameters, such as the luminosity distance, will be trivial and will
not need to be computed numerically.)

As an example, consider a binary with two non-spinning $10\,M_\odot$
black holes at a distance of $200\,{\rm Mpc}$.  The inspiral SNR for
this system is $\sim 100$ for advanced LIGO interferometers
\cite{paperI}.  In this optimistic case, the information obtained from
the inspiral waveform will be roughly as follows \cite{note2}: The
distance to the system will be known to $\alt 2 \%$, the masses will
be known to $\sim 40 \%$ (although the chirp mass ${\cal M} = 
{\mu}^{3/5} M^{2/5}$ will likely be known to an accuracy of $\alt
0.1\%
$), the arrival time to $\sim 0.1 \, {\rm ms}$, the position on
the sky to less than one square degree, and the angles defining ${\hat
{\bf L}}$ and $\phi_c$ to $\alt 10^\circ$.  Also, some information
will be obtained about two particular combinations of the spins ${\bf
S_1}$ and ${\bf S_2}$ (see
Refs.~\cite{cutlerflan94,krolaknew,poissonwill} for details).  As a
second example, consider a binary of two $15\,M_\odot$ black holes at
$z=1$, for which the inspiral SNR for advanced interferometers is
$\sim 7$ \cite{paperI}.  For such a binary the accuracies are several
times worse.  The luminosity distance is measured to $\sim 20\,\%$,
for example, and although the chirp mass is measured to $\alt 1 \%$,
the individual masses are only constrained to lie in the ranges $3
\,M_\odot \alt m_2 \alt 15 \,M_\odot$ and $15
\,M_\odot \alt m_1 \alt 100 \,M_\odot$ \cite{note2}.

Turn, now, to the information obtainable from inspiral signals for the
space-based LISA interferometer.  Equation (A6) of Ref.~\cite{paperI}
shows that the time $T_{\rm insp}$ which the gravitational wave signal
spends in the interferometer's bandwidth during the inspiral before
merger is approximately  
\begin{equation}
T_{\rm insp} \sim 0.4 \, {\rm yr} \, \left[ {(1+z) M \over 10^6 M_\odot}
\right]^{-5/3} \, \left[ 1 - \left({ (1+z) M \over 4 \times 10^7 \,
M_\odot}\right)^{8/3} \right].
\end{equation}
Signal-to-noise ratios from such inspirals (or from the last year of
inspiral if $T_{\rm insp} \ge 1 \, {\rm yr}$) will be $\agt 100$ for
all events with cosmological redshift $z \alt 10$ and with $10^4
M_\odot \alt (1+z) M \alt 5 \times 10^7 M_\odot$; see Fig.\ 6 of Ref.\
\cite{paperI}.  Thus, detailed information about the binary's
parameters should be available for analyzing merger signals detected
by LISA \cite{LISAacc}.  (For some LISA BBH sources, most of the
inspiral SNR will 
come from the IBBH regime $6M \alt r \alt 12 M$ discussed in the
Introduction.  For such sources, accurate IBBH templates will likely
be needed to extract all the available inspiral information.)

In some cases with LIGO/VIRGO, and in many cases with LISA, it will
also be possible to analyze the ringdown waveform using optimal
filtering to extract the ringdown frequency and damping time
\cite{finnmeasure,echeverria}.  These measurements will yield the
mass $M$ and spin parameter $a$ of the final black hole.  The
accuracy of such measurements will be approximately given by
\cite{finnmeasure,echeverria}
\begin{eqnarray}
{\Delta M \over M} &\simeq & {2 (1 - a)^{9/20} \, \over (S/N)_{\rm
ringdown}} \nonumber \\
\Delta a & \simeq & {6 (1 - a)^{1.06} \, \over (S/N)_{\rm
ringdown}},
\end{eqnarray}
where $(S/N)_{\rm ringdown}$ is the matched filtering SNR for the
ringdown signal.  It should also be possible to measure the time at
which the ringdown starts to within an accuracy $ \alt 1 / f_{\rm
qnr}$.  For low mass coalescences ($M \alt 50 M_\odot$), such
measurements will only be possible for the very strongest detected
events: the ringdown SNR will be $\agt 1$ only for the strongest $\sim
1\%$ of detected events for initial and advanced LIGO interferometers
\cite{ringnote2}.  For larger mass BBH coalescences, however, the
ringdown SNR will be larger, as can be seen from Figs.~4 and 5 of
Ref.~\cite{paperI}, and ringdown measurements will be feasible for a
reasonable fraction of detected signals.  For LISA, Fig.~6 of
Ref.~\cite{paperI} shows that most detected merger events will be
accompanied by easily detectable ringdown signals with SNR values
$\agt 100$.  Thus, accurate values of $M$ and $a$ should be available
as prior information when analyzing merger signals detected by LISA.

For the strongest detected signals, it may also be possible to measure
the complex amplitudes of some of the quasinormal modes in the
waveform other than the dominant $l=m=2$ mode.  These higher order
quasinormal ringing (QNR) modes will not be as long lived as the
$l=m=2$ mode, but they may 
nevertheless be detectable.  The amplitudes and phases of such modes
will constitute very useful information if they are measurable, since
their values should be predicted by the supercomputer simulations as
functions of the binary's parameters at the start of the merger phase.
The supercomputer simulations will have passed an important test if
the measured mode amplitude values are consistent with known
information about the initial conditions.

\section{ANALYSIS OF THE MERGER WAVES WITHOUT
TEMPLATES---VISIBILITY OF MERGER SIGNAL AFTER
BAND-PASS FILTERING}
\label{visibility}

Turn now to the data analysis of the merger waves, focusing
on the case in which matched filtering cannot be used.  This situation
will arise if supercomputer simulations are unable to produce merger
templates, or if they have only produced a small sampling of the total
function space ${\cal S}$ of merger waveforms when BBH signals are detected.
Such a sampling should provide valuable qualitative information about
the merger waveforms (as we discuss in Sec.~\ref{notmplsuper} below),
but would be too sparse to be used as a bank of optimal filters.
(As mentioned in Sec.~\ref{extractionsec} above, it may be possible to
perform matched filtering in the absence of a complete set of
templates, but this is not guaranteed).

In the absence of a complete set of theoretical templates, one's first
aim will be to reconstruct from the noisy detector output a best-guess
estimate of the merger waveform $h(t)$ \cite{noteF}.  If a small
number of representative supercomputer templates are available, it may
then be possible to interpret the measured waveform and obtain
qualitative information about the BBH source.  One very simple
procedure that could be used to obtain an estimate of the waveform
shape is simply to band-pass filter the data stream according to our
prior prejudice about the frequency band of the merger waves (based on
estimates of the merger signal bandwidth \cite{paperI}, hopefully
supplemented by information from representative supercomputer
simulations and from inspiral/ringdown measurements) \cite{note20}.
However, after such band-pass filtering, the merger signal may be
dominated by detector noise and may not even be visible.  (Signals
that are visible in the noise will clearly be easier to reconstruct
from the noisy data stream; we demonstrate this mathematically in
Sec.\ \ref{fidelity} below).

In this section we explore this issue of merger waveform 
visibility, by which we mean whether or not the signal stands out
above the noise after band-pass filtering.
A signal will be visible if the band-pass filtering
SNR is large compared to unity; see, for example, the discussion in
Ref.~\cite{paperI}.  We use the results of Ref.~\cite{paperI} to
estimate band-pass filtering SNRs, first for the inspiral waves near
the end of the inspiral in Sec.~\ref{inspiralvisibility}, and then for
the merger waves in Sec.~\ref{mergervisibility}.  
The analysis of the inspiral waves is useful as background for the merger
visibility calculation, and is also indicative of the visibility
of the early merger waves (if the endpoint of inspiral is visible with
band-pass filters, than one would expect that by continuity the beginning
of the merger should be visible as well).

\subsection{Visibility of inspiral waveform}
\label{inspiralvisibility}

We focus on BBH events which have been detected via their inspiral
waves using matched filtering.  Since the event has been detected, the
inspiral matched filtering SNR must be $\agt 6$ \cite{prl}; however,
it does not follow that the inspiral signal is visible in the data
stream without matched filtering.  (In fact, for neutron star-neutron
star binaries the reverse is usually the case: the amplitude of the
signal is rather less than the noise, and so matched filtering is very
necessary to detect the waves.)  We now estimate 
the degree of visibility of the last few cycles of the inspiral
waveform for BBH coalescences.

The dominant harmonic of the inspiral waveform can be written as
\begin{equation}
h(t) = h_{\rm amp}(t) \cos [ \Phi(t)],
\end{equation}
where the amplitude $h_{\rm amp}(t)$ and instantaneous frequency
$f(t)$ 
[given by $2 \pi f(t) = d \Phi / d t$] are slowly evolving.  For such
waveforms, the SNR squared obtained using band-pass filtering is
approximately given by the SNR squared {\it per cycle} obtained from
matched filtering [{\it cf.}\ Eq.\ (2.9) of Ref.~\cite{paperI}]:
\FL
\begin{eqnarray}
\left({S \over N}\right)^2_{\rm band-pass} &\approx& \left(S\over
N\right)^2_{\rm optimal\ filter,\ per\ cycle} \nonumber \\ &=& \,
\left[{h_{\rm amp}\left[t(f)\right] \over h_n(f)}\right]^2.
\label{snrpcinspiral}
\end{eqnarray}
In Eq.\ (\ref{snrpcinspiral}), an rms average over source orientations
has been performed, $t(f)$ denotes the time at which the instantaneous
frequency has value $f$, and $h_n(f) \equiv \sqrt{5 f S_h(f)}$.
Note that the band-pass filtering SNR (\ref{snrpcinspiral}) is
evaluated at a specific frequency, whereas typically when one
discusses matched filtering SNRs, an integral over a large frequency
band has been performed.  Next, we insert the value
of $h_{\rm amp}[t(f)]^2$ for the leading-order approximation to the
inspiral waves, which can be obtained from, for example, Eq.~(3.20) of
Ref.\ \cite{paperI}, and obtain 
\begin{equation}
\left({S \over N}\right)^2_{\rm band-pass} 
= {64 \pi^{4/3} {\cal M}^{10/3} (1+z)^{10/3} f^{4/3} \over 5
D(z)^2 h_n(f)^2}.
\label{snrpcinspiral1}
\end{equation}
Here ${\cal M}\equiv \mu^{3/5} M^{2/5}$ is the chirp mass, $z$ is the
binary's cosmological redshift and $D(z)$ is the binary's luminosity
distance.

In Eq.\ (4.1) of Ref.\ \cite{paperI} we introduced an analytic formula
for a detector's noise spectrum $S_h(f)$, which, by specialization of
its parameters, could describe to a good approximation either an
initial LIGO interferometer, an advanced LIGO interferometer, or a
space-based LISA interferometer.  We now insert that formula into Eq.\
(\ref{snrpcinspiral}), and specialize to the frequency
\begin{equation}
f = f_{\rm merge} = {\gamma_m \over (1+z) M},
\label{fmergedef}
\end{equation}
where $\gamma_m = 0.02$.  The frequency $f_{\rm merge}$ is approximately the
location of the transition from inspiral to merger, as estimated
in Ref.\ \cite{paperI}.  We thus obtain for the band-pass filtering
SNR 
\begin{equation}
\left({S\over N}\right)^2_{\rm band-pass} \approx { 4 \pi^{4/3}
M^5 (1+z)^5 \gamma_m^{-5/3} \alpha^3 f_m^3 \over 5 D(z)^2 h_m^2},
\label{snrpcinsp}
\end{equation}
where $\alpha$, $h_m$ and $f_m$ are the parameters used in
Ref.\ \cite{paperI} to describe the interferometer noise curve.
Equation (\ref{snrpcinsp}) is valid only when the redshifted mass $(1+z) M$
of the binary is smaller than $\gamma_m /\alpha f_m$.

For initial LIGO interferometers, appropriate values of the parameters
$h_m$, $f_m$ and $\alpha$ are given in Eq.~(4.2) of Ref.\
\cite{paperI}.  Inserting these values into Eq.~(\ref{snrpcinsp}) gives
\begin{equation}
\left( {S \over N} \right)_{\rm band-pass} \sim \, 1.1 \, \left[ {200
\, {\rm Mpc} \over D(z) }\right] \, \left[ { (1 + z) M \over 20 \,
M_\odot } \right]^{5/2},
\label{sbd3}
\end{equation}
which is valid for $(1+z) M \alt 18 M_\odot$.  Now, the SNR obtained
by matched filtering the inspiral signal ({\it i.e.}, by correlating the
inspiral data with an inspiral template over the full bandwidth of the
signal) is approximately \cite{paperI}
\begin{equation}
\left( {S \over N} \right)_{\rm optimal} \sim \, 2.6 \, \left[ {200
\, {\rm Mpc} \over D(z) }\right] \, \left[ { (1 + z) M \over 20 \,
M_\odot } \right]^{5/6}.
\label{snroptinsp}
\end{equation}
Also the quantity (\ref{snroptinsp}) must be $\agt 6$ \cite{prl}, 
because, by assumption, the inspiral has in fact been detected.
By eliminating the luminosity distance $D(z)$ between Eqs.~(\ref{sbd3})
and (\ref{snroptinsp}) we find that the band-pass filtering SNR for
the last inspiral cycles of detected binaries satisfies
\begin{equation}
\left( {S \over N} \right)_{\rm band-pass} \agt \, 2.5 \,\left[ { (1 +
z) M \over 20 \, M_\odot } \right]^{5/3}.
\label{sbd4}
\end{equation}
Therefore, the last few cycles of the inspiral should be individually
visible above the noise for BBH events with $5 M_\odot \alt M \alt 20
M_\odot$ detected by initial LIGO interferometers.

We now repeat the above calculation with the values of $h_m$, $f_m$, and
$\alpha$ appropriate for advanced LIGO interferometers, which are
given in Eq.\ (4.3) of Ref.\ \cite{paperI}.  The band-pass filtering
SNR for advanced interferometers is
\begin{equation}
\left( {S \over N} \right)_{\rm band-pass} \sim \, 1.6 \, \left[ {1
\, {\rm Gpc} \over D(z) }\right] \, \left[ { (1 + z) M \over 20 \,
M_\odot } \right]^{5/2},
\label{sbd5}
\end{equation}
and the SNR obtained by matched filtering the inspiral signal is
\begin{equation}
\left( {S \over N} \right)_{\rm optimal} \sim \, 16 \, \left[ {1
\, {\rm Gpc} \over D(z) }\right] \, \left[ { (1 + z) M \over 20 \,
M_\odot } \right]^{5/6},
\label{snroptinspA}
\end{equation}
for $(1+z) M \alt 37 M_\odot$ \cite{paperI}.  So, with the assumption
that $(S/N)_{\rm optimal} \agt 6$, we 
find
\begin{equation}
\left( {S \over N} \right)_{\rm band-pass} \agt \, 0.6 \,\left[ { (1 +
z) M \over 20 \, M_\odot } \right]^{5/3}
\end{equation}
for $(1+z) M \alt 37 M_\odot$.  Therefore, for BBH inspirals with
$(1+z) M \alt 37 M_\odot$ detected by advanced LIGO interferometers,
the last few cycles of the inspiral will be just barely individually
visible above the noise, depending on the binary's total mass $M$.
The last few cycles of the inspiral will also be visible for larger mass BBH
systems, as can be seen by combining Eq.~(\ref{snrpcinspiral}) above
with Figs.~4 and 5 of Ref.~\cite{paperI}.

For LISA, Eq.~(\ref{snrpcinsp}) combined with Eq.~(4.3) of
Ref.~\cite{paperI} yields
\begin{equation}
\left( {S \over N} \right)_{\rm band-pass} \sim \, 180 \, \left[ {1
\, {\rm Gpc} \over D(z) }\right] \, \left[ { (1 + z) M \over 10^6 \,
M_\odot } \right]^{5/2}
\end{equation}
for $(1+z) M \alt 10^5 \, M_\odot$, with larger values for
$10^5 \, M_\odot \alt (1+z) M \alt 3 \times 10^7 \,M_\odot$.  Therefore
individual cycles of the inspiral waveform should be clearly visible
for LISA. 

\subsection{Visibility of merger waveform}
\label{mergervisibility}

Consider now the merger waveform itself.  This will be visible if the
SNR from band-pass filtering of the merger signal is large compared to
unity.  In Ref.~\cite{paperI} we showed that
\FL
\begin{equation}
\left( { S\over N} \right)_{\rm band-pass,merger} \approx {1 \over \sqrt{{\cal
N}_{\rm bins}} } \, \left( {S \over N} \right)_{\rm optimal,merger},
\label{snrbp}
\end{equation}
where ${\cal N}_{\rm bins} = 2 T \Delta f$; $T$ and $\Delta f$ are
the expected duration and bandwidth of the merger signal.  We also
estimated [Eq.~(3.32) of Ref.~\cite{paperI}] that for the merger
waves, 
\begin{equation}
\sqrt{{\cal N}_{\rm bins}} \sim 5,
\label{eq332}
\end{equation}
although there is large uncertainty in this estimate and ${\cal
N}_{\rm bins}$ will vary from event to event.  Combining Eqs.\ (5.4)
of Ref.\
\cite{paperI} for initial LIGO interferometers, Eq.\ (\ref{eq332}),
and the threshold for detection
\cite{prl} 
\begin{equation}
\left( {S \over N} \right)_{\rm optimal,\,inspiral} \agt 6
\end{equation}
yields
\begin{equation}
\left( { S\over N} \right)_{\rm band-pass,\, merger} \agt 0.8
\left[ {(1+z) M \over 20 M_\odot}\right]^{5/3}
\label{initSNbp}
\end{equation}
for $(1+z) M \alt 18 M_\odot$.  For advanced LIGO interferometers,
Eq.~(\ref{eq332}) together with Eq.~(5.5) of Ref.~\cite{paperI}
similarly yield 
\begin{equation}
\left( { S\over N} \right)_{\rm band-pass,\, merger} \agt 0.2
\left[ {(1+z) M \over 20 M_\odot}\right]^{5/3}
\label{advSNbp}
\end{equation}
for $(1+z) M \alt 37 M_\odot$.  Note that, contrary to one's
intuition, the value (\ref{advSNbp}) for advanced interferometers is
lower than the value (\ref{initSNbp}) for initial interferometers.
This is because the advanced interferometers can detect inspirals with
lower band-pass filtering SNRs than the initial interferometers, due
to the larger number of cycles of the inspiral signal in the advanced
interferometer's bandwidth.  Matched filtering is extremely efficient
at detecting inspiral signals, and it is more so for advanced
interferometers than for initial interferometers.  The weaker the
signals that are detectable by matched filtering, the less visible the
merger waveform will be after bandpass filtering.

The SNR values (\ref{initSNbp}) and (\ref{advSNbp}) indicate that for
typical inspiral-detected BBH systems with $M\alt 20 M_\odot$ (initial
interferometers) or $M\alt 40M_\odot$ (advanced interferometers), the
merger signal will not be easily visible in the noise, and that only
the somewhat rarer, closer events will have easily visible merger
signals.  This conclusion is somewhat tentative because of the
uncertainty in the estimates of ${\cal N}_{\rm bins}$ and of the
energy spectra discussed in Ref.~\cite{paperI}.  Also the visibility
of the merger waveform will probably vary considerably from event to
event.

This conclusion only applies to low mass BBH systems which are
detected via their inspiral waves.  For higher mass systems which are
detected directly via their merger and/or ringdown waves, the merger
signal should be visible above the noise after appropriate band-pass
filtering.  Moreover, most merger events detected by LISA will have
band-pass filtering SNRs $\gg 1$, as can be seen from Fig.~6 of Ref.\
\cite{paperI}, and thus should be easily visible.

Our crude visibility argument thus suggests that the prospects
for accurately recovering the merger waveform are good only for the
stronger detected merger signals.  This visibility analysis also
illustrates the importance of theoretical template waveforms: the SNRs
that can be achieved without them will often be mediocre at best.
Templates for the merger will be able to boost measured SNRs by a
factor $\sqrt{{\cal N}_{\rm bins}} \sim 5$.
Of course, we need to go beyond this simple analysis and try to
determine the optimal method of reconstructing the shape of the merger
waveform from the noisy data; we propose one method in the following
section.

\section{ANALYSIS OF THE MERGER WAVES WITHOUT
TEMPLATES---A METHOD OF EXTRACTING A BEST-GUESS
MERGER WAVEFORM FROM THE NOISY DATA STREAM}
\label{bestguess}

\subsection{Overview}
\label{bestguessoverview}

In the absence of a complete set of theoretical templates we would
like to reconstruct from the noisy detector data stream a best-guess
estimate of the merger waveform $h(t)$.  In this section, we suggest
and describe
a method, based on the technique of maximum likelihood estimation
\cite{wainandzub,helstrom}, for performing such a waveform
reconstruction.

A method for estimating the merger waveform shape $h(t)$
should use all available prior knowledge about the waveform.  We will
hopefully know 
from representative supercomputer simulations and perhaps from the
measured inspiral/ringdown signals the following: (i) the approximate
starting time of the merger; (ii) the fact that it starts off strongly
(smoothly joining on to the inspiral waveform) and eventually dies
away in quasinormal ringing; and (iii) the approximate bandwidth and
duration of the signal.  For those signals for which both the inspiral
and the ringdown are strong enough to be detectable with optimal
filtering, the duration of the merger portion of the waveform will be
fairly well known, as will the frequency $f_{\rm qnr}$ of the ringdown
signal onto which the merger waveform must smoothly join.  The
technique which we describe in this section encodes such prior
information and makes use of it in reconstructing the best-guess
estimate of the waveform.

We shall describe this method in the context of a single detector or
interferometer.  However, in a few years there will be in operation a
network of several detectors (both interferometers
\cite{ligoscience,virgo,geo} and resonant mass antennae) and
from the combined outputs of these several detectors one would like to
reconstruct the two independent polarization components $h_+(t)$ and
$h_\times(t)$ of the gravitational waves from the merger.  In Appendix
\ref{network} we show how to extend the waveform estimation method
discussed in this section to an arbitrary number of detectors, which
yields a method of reconstructing the two waveforms $h_+(t)$ and
$h_\times(t)$.

The issue of reconstructing the waveforms $h_+(t)$ and $h_\times(t)$
was previously addressed by G\"ursel and Tinto \cite{tinto}, in the
context of a network of three interferometers and for arbitrary bursts
of gravitational waves.  G\"ursel and Tinto suggest a method of
extracting, from the outputs of all the interferometers (i) the
direction to the source, and (ii) the two gravitational waveforms.
For many BBH mergers, the direction to the source will have already
been determined to fairly good accuracy from the inspiral waveform
\cite{krolakxxx}, and so the G\"ursel-Tinto filtering method is not
directly applicable.  However, they do suggest in passing a method for
extracting the waveforms $h_+(t)$ and $h_\times(t)$ when the direction
to the source is given.  In Appendix \ref{network} we show that
our filtering method (as extended to a network of interferometers) is
an extension and generalization of the G\"ursel-Tinto algorithm.

The filtering methods which we consider are based on the theory of maximum
likelihood estimation {\cite{wainandzub,helstrom}}.  The use of maximum
likelihood estimators has been discussed extensively by many authors in
the context of gravitational waves of a known functional form,
depending only on a few parameters
{\cite{finn2,cutlerflan94,krolaknew,krolakxxx,montecarlo}}.  In this 
section we consider their application to gravitational wave bursts of
largely unknown shape.   The resulting data analysis methods which we
derive are closely related mathematically to the methods discussed
previously {\cite{finn2,cutlerflan94,krolaknew,krolakxxx,montecarlo}},
but are considerably different in operational terms and in
implementation.

\subsection{Derivation of data analysis method}
\label{deriveBG}

We now turn to our derivation of the best-guess waveform estimator
using maximum likelihood estimation.  Suppose that our prior
information about the merger waves includes the information that they
lie inside some time interval of duration $T$, and inside some
frequency interval of length $\Delta f$.  We define ${\cal N}_{\rm
bins} = 2 T \Delta f$, {\it cf}. Sec.~\ref{notations} above.  We also
suppose that we have a stretch of data to analyze of duration
$T^\prime > T$ and with sampling time $\Delta t < 1 / (2 \Delta f)$.
These data lie in a linear space $V$ of dimension
\begin{equation}
{\cal N}_{\rm bins}^\prime = 2 T^\prime / \Delta t
\end{equation}
which is strictly larger than ${\cal N}_{\rm bins}$.  Thus, ${\cal
N}_{\rm bins}^\prime$ is the number of independent real data points in
the data, and ${\cal N}_{\rm bins}$ is the number of independent real
data points in that subset of the data which we expect to contain the
merger signal.  Note that these definitions constitute a
modification/extension of the conventions introduced in
Sec.~\ref{notations} above, where the dimension of the space $V$ was
denoted by ${\cal N}_{\rm bins}$.  We will use, unmodified, the other
conventions of Sec.~\ref{notations}: thus, the detector output ${\bf
s}$ is given by ${\bf s} = {\bf h} + {\bf n}$, where ${\bf h}$ the
gravitational-wave signal and ${\bf n}$ the detector noise, and the
vectors ${\bf s}$, ${\bf h}$ and ${\bf n}$ are all elements of the
vector space $V$ of dimension ${\cal N}_{\rm bins}^\prime$.

In our analysis below, we will allow the basis of the vector space $V$
to be arbitrary.  Thus, $n_i$ (for example) will denote the components
of the noise on this arbitrary basis.  However, we will occasionally
specialize to the time-domain and frequency-domain bases discussed in
Sec.~\ref{notations} above.  We will also consider wavelet bases of
$V$.  Wavelet bases can be regarded as any set of functions
$w_{ij}(t)$ such that $w_{ij}(t)$ is approximately localized in time
at the time $t_i = t_{\rm start} + (i/n_T) T^\prime$, and
approximately localized in frequency at the frequency $f_j = (j/n_F)
(\Delta t)^{-1}$.  The index $i$ runs from 1 to $n_T$ and $j$ from
$-(n_F-1)/2$ to $(n_F-1)/2$.  Clearly the number of frequency bins
$n_F$ and the number of time bins $n_T$ must satisfy $n_T n_F = {\cal
N}_{\rm bins}^\prime$, but otherwise they can be arbitrary; typically
$n_T \sim n_F \sim \sqrt{{\cal N}_{\rm bins}^\prime}$.  Also, the
functions $w_{ij}$ usually all have the same shape, so that
\begin{equation}
w_{ij}(t) \propto \varphi\left[ f_j(t - t_i) \right],
\label{sameshape}
\end{equation}
for some function $\varphi$.  For our considerations here, the shape
of $\varphi$ is not of critical importance.  Also, wavelet bases are
often overcomplete; the bases we discuss below are to be considered
simply complete.  So if the full function space of some family of
wavelets is $W$, we restrict ourselves to some complete subset $W'$ of
that space.  The advantage of wavelet bases is that they they
simultaneously encode frequency domain and time domain information.

Let $p^{(0)}({\bf h})$ be the probability distribution (PDF) that
summarizes our prior information about the gravitational waveform.  A
standard Bayesian analysis shows that the PDF of ${\bf h}$ given the
measured data stream ${\bf s}$ is
\cite{finn2,finnmeasure}
\FL
\begin{equation}
p({\bf h}\,|\,{\bf s}) = {\cal K} \, p^{(0)}({\bf h}) \exp\left[ -
\Gamma_{ij} (h^i-s^i) (h^j - s^j)/2\right],
\label{pdf0}
\end{equation}
where the matrix $\Gamma_{ij}$ is defined in Eq.~(\ref{Gammadef}) and
${\cal K}$ is a normalization constant {\cite{finn2}}.  In principle
this PDF gives complete information about the measurement.  Maximizing
the PDF will yield the maximum likelihood estimator for the merger
waveform ${\bf h}$.  This estimator will be some function ${\bf h} =
{\bf h}({\bf s})$, which in general will be a non-linear function.  
The effectiveness of the resulting
estimator of the waveform will depend on how much prior information
concerning the waveform shape can be encoded in the choice of prior
PDF $p^{(0)}$.

One of the simplest possibilities is to take $p^{(0)}$ to be
concentrated on some linear subspace $U$ of the space $V$, and to be
approximately constant inside this subspace.  A multivariate Gaussian
with widths very small in some directions and very broad in others
would accomplish this to a good approximation.  For such choices of
prior PDF $p^{(0)}$, the resulting maximum likelihood estimator [the
function ${\bf h} = {\bf h}({\bf s})$ that maximizes the PDF
(\ref{pdf0})] is simply the perpendicular projection $P_U$ of ${\bf
s}$ into $U$:
\begin{equation}
{\bf h}_{\rm best-fit}({\bf s}) = P_U({\bf s}),
\label{bestfit}
\end{equation}
where
\begin{equation}
P_U({\bf s}) \equiv \sum_{i,j=1}^{n_U} \, u^{ij} \left({\bf u}_j\,|\,{\bf
s}\right) \,\, {\bf u}_i.
\label{projection}
\end{equation}
Here, ${\bf u}_1, \ldots, {\bf u}_{n_U}$ is an arbitrary basis of $U$,
$n_U$ is the dimension of $U$, $u^{ij} u_{jk} = \delta^i_j$ and
$u_{jk} = \left({\bf u}_j\,|\,{\bf u}_k\right)$.

We remark that the method of filtering (\ref{bestfit}) is a special
case of Wiener optimal filtering: it is equivalent to optimal
filtering with templates that are constructed by taking linear
combinations of the basis functions
${\bf u}_i$.  (The equivalence between maximum likelihood estimation
and Wiener optimal filtering in more general contexts has been shown
by Echeverria {\cite{Fernando}}.)  To show that our filtering method
is a form of Wiener optimal filtering, define a family of template
waveforms that depends on parameters $a_1,\ldots, a_{n_U}$ by
\begin{equation}
h(t;a_j) = \sum_{j=1}^{n_U} \, a_j u_j(t),
\label{linearfamily}
\end{equation}
where $u_j(t)$ are the functions of time corresponding to the basis
elements ${\bf u}_j$ of $U$.  If $s(t)$ is the measured detector
output, define for any function $h(t)$
\begin{equation}
{S\over N}\left[h(t)\right] \equiv { \left( {\bf h} \, | \, {\bf s} \right) \over
\sqrt{ \left( {\bf h} \, | \, {\bf h} \right)} }.
\label{snrgeneral1}
\end{equation}
This is the SNR for the template $h(t)$ with the data stream {\bf s}.
The best-fit signal given by the optimal filtering method is the
template which maximizes the SNR
(\ref{snrgeneral1}), {\it i.e.}, the template $h(t;{\hat a}_j)$ such
that
\begin{equation}
{S\over N}\left[h(t;{\hat a}_j)\right] = \max_{a_1,\ldots,a_{n_U}} \,
{S\over N}\left[h(t; a_j)\right].
\end{equation}
However, it is easy to show from
Eqs.~(\ref{projection})--(\ref{snrgeneral1}) that
\begin{equation}
P_U({\bf s}) = h(t;{\hat a}_j).
\label{bestfits}
\end{equation}
Thus, computing the perpendicular projection (\ref{projection}) of
${\bf s}$ into $U$ is equivalent to Wiener optimal filtering with the
family of templates (\ref{linearfamily}).  From an operational point
of view, the method of filtering (\ref{projection}) is quite different
to the normal implementation of optimal filtering, which is carried
out by calculating the SNR (\ref{snrgeneral1}) for various parameter
values, but the final best-fit signals (\ref{bestfits}) are identical.
[Of course, Wiener optimal filtering is normally only carried out when
the dependence of the waveform $h(t;a_j)$ on the parameters $a_j$ is
complicated and nonlinear, as when searching for inspiral waves
where the parameters represent astrophysical characteristics of
the binary system.]

To summarize, the maximum likelihood estimator (\ref{bestfit}) gives a
general procedure for specifying a filtering algorithm adapted to a
given linear subspace $U$ of the space of signals $V$.  We will
suggest below a specific choice for the subspace $U$; but first, we discuss
some general issues related to making such a choice.

At the very least, we would like our choice of $U$ to effect
truncation of the measured data stream in both the time domain and the
frequency domain, down to the intervals of time and frequency in which
we expect the merger waveform to lie.  (We assume that the duration of
the data being analyzed, $T^\prime$, will be somewhat longer than
one's guess of the merger duration, $T$.)  Because of the uncertainty
principle, such a truncation cannot be done exactly.  Moreover, for
fixed specific intervals of time and of frequency, there are
different, inequivalent ways of approximately truncating the signal to
these intervals {\cite{eg}}.  The differences between the inequivalent
methods are essentially due to aliasing effects.  Such effects cannot
always be neglected in the analysis of merger waveforms, because the
duration $T
\sim 10 M$ -- $100 M$ \cite{paperI} of the waveform is probably only a
few times larger than  the reciprocal of the highest frequency of
interest. 

It turns out that the simplest method of truncating in frequency
(band-pass filtering) is, to a good approximation, a projection of the
type (\ref{bestfit}) that we are considering.  Truncating in the time
domain, on the other hand, is not a projection of this type.

Let us first discuss band-pass filtering.  Let ${\bf d}_k$ [{\it cf}.\
Eq.\ (\ref{freqbasis})] be a frequency domain basis of $V$.  For a
given frequency interval $\left[f_{\rm char} - \Delta f/2,f_{\rm
char}+\Delta f/2\right]$, let $U$ be the subspace of $V$ spanned by
the elements ${\bf d}_j$ with $|f_{\rm char} - f_j| < \Delta f /2$,
{\it i.e.}, the 
span of the subset of the frequency domain basis that corresponds to
the given frequency interval.  Then the projection operation $P_U$ is
to a moderate approximation just the band-pass filter:
\begin{equation}
P_U\left[ \sum_{j=1}^{{\cal N}_{\rm bins}^\prime} \,s^j \, {\bf
	d}_j\right] \approx {\sum}^\prime \,s^j\, {\bf d}_j,
\label{freqtruncok}
\end{equation}
where the notation $\sum^\prime$ means that the sum is taken only over
the appropriate 
range of frequencies.  The reason for the relation (\ref{freqtruncok})
is that the basis ${\bf d}_j$ is approximately orthogonal with respect
to the noise inner product (\ref{discreteinnerprod}): different
frequency components of the noise are statistically independent up to
small aliasing corrections of the order of $\sim 1/(f_{\rm char}
T^\prime)$.  Thus, if our {\it a priori\/} information is that
the signal lies within a certain frequency interval, then the maximum
likelihood estimate of the signal is approximately given by passing
the data stream through a band-pass filter.

An analogous statement is not true in the time domain.  If our {\it a
priori} information is that the signal vanishes outside a certain
interval of time, then truncating the data stream by throwing away the
data outside of this interval will not give the maximum likelihood
estimate of the signal.  This is because of statistical correlations
between sample points just inside and just outside of the time
interval: the measured data stream outside the interval gives
information about what the noise inside the interval is likely to be.
These correlation effects become unimportant in the limit $T f_{\rm
char} \to \infty$,  but for BBH merger signals $T f_{\rm char}$
is probably $ \alt 20$ \cite{paperI}.  The
correct maximum likelihood estimator of the waveform, when our prior
information is that the signal vanishes outside of a certain time
interval, is given by Eq.~(\ref{projection}) with the basis $\{{\bf
u}_1, \ldots, {\bf u}_{n_U}\}$ replaced by the appropriate subset of
the time-domain basis $\{{\bf e}_1, \ldots, {\bf e}_{{\cal N}_{\rm
bins}^\prime}\}$ discussed in Sec.~\ref{notations}.

Our suggested choice of subspace $U$ and corresponding specification
of a filtering method is as follows.  Pick a wavelet basis ${\bf
w}_{ij}$ of the type discussed above.  (The filtering method will
depend only weakly on which wavelet basis is chosen).  Then, the
subspace $U$ is taken to be the span of a suitable subset of this
wavelet basis, according to our prior prejudice regarding the
bandwidth and duration of the signal.  The dimension
$n_U$ of $U$ will be given by
\begin{equation}
n_U = {\cal N}_{\rm bins} = 2 T \Delta f.
\end{equation}

In more detail, the filtering method would work as follows.  First,
band-pass filter the data stream and truncate it in time, down to
intervals of frequency and time that are several times larger than are
ultimately required, in order to reduce the number of independent data
points ${\cal N}_{\rm bins}^\prime$ to a manageable number.  Second,
for the wavelet 
basis ${\bf w}_{ij}$ of this reduced data set, calculate the matrix
$w_{ij\,i^\prime j^\prime} = \left({\bf w}_{ij}\,|{\bf w}_{i^\prime
j^\prime}\right)$.  Recall that the index $i$ corresponds to a time
$t_i$, and the index $j$ to a frequency $f_j$ [cf.\ the discussion
preceding Eq.~(\ref{sameshape})].  Third, pick out the sub-block
${\bar w}_{ij\,i^\prime j^\prime}$ of the matrix $w_{ij\,i^\prime
j^\prime}$ for which the times $t_i$ and $t_{i^\prime}$ lie in the
required time interval, and for which the frequencies $f_j$ and
$f_{j^\prime}$ lie in the required frequency interval.  Numerically
invert this matrix to obtain ${\bar w}^{ij\,i^\prime j^\prime}$.
Finally, the best-fit waveform is given by
\begin{equation}
{\bf h}_{\rm best-fit} = {\sum_{ij}}^\prime
{\sum_{{i^\prime}{j^\prime}}}^\prime \ {\bar w}^{ij\,i^\prime j^\prime}
\, \left( {\bf s} \, | \, {\bf w}_{i^\prime j^\prime} \right) \, {\bf
w}_{ij}, 
\label{bestfit2}
\end{equation}
where $\sum^\prime$ means the sum over the required time and frequency
intervals. 

Note that the best-fit signal in this case is {\it not} given by first
taking the finite wavelet transform of the reduced data ({\it i.e.},
finding the coefficients $s^{ij}$ in the expansion ${\bf s} =
\sum_{ij} s^{ij} {\bf w}_{ij}$) and then throwing away the
coefficients outside of the required time and frequency intervals,
which would yield
\begin{equation}
{\sum_{ij}}^\prime \,s^{ij} \,{\bf w}_{ij}.
\end{equation}
Note also that the best fit signal (\ref{bestfit2}) would also be
obtained by 
calculating the SNR (\ref{snrgeneral1}) for the family of waveforms
\begin{equation}
h(t) = {\sum_{ij}}^\prime \ c^{ij} w_{ij}(t)
\end{equation}
and by maximizing over the $c_{ij}$'s, as discussed above.  This
essentially corresponds to building a family of templates with the
wavelet basis, and then performing matched filtering with that bank of
templates.

\subsection{Extension of method to incorporate other types of prior
information} 
\label{extendedmethod}

A more sophisticated filtering method can be
obtained by a generalization of the above analysis.  Let us suppose
that the prior PDF $p^{(0)}({\bf h})$ is a general multivariate
Gaussian in ${\bf h}$.  For example, one could choose the prior PDF to
be of the form
\begin{equation}
p^{(0)}({\bf h}) \, \propto \, \exp \left[ - {1 \over 2} \sum_{ij} \, {
(h_{ij} - {\bar h}_{ij})^2 \over \alpha_{ij}^2 } \right],
\end{equation}
where $h^{ij}$ are the expansion coefficients of the signal ${\bf h}$
on some fixed wavelet basis ${\bf w}_{ij}$, so that ${\bf h} =
\sum_{ij} h^{ij} {\bf w}_{ij}$.  Then, by making suitable
choices of the parameters ${\bar h}_{ij}$ and $\alpha_{ij}$, such a
PDF could be chosen to encode the information that the frequency content
of the signal at early times is concentrated near $f_{\rm merge}$,
that the signal joins smoothly onto the inspiral waveform, that at the
end of merger the dominant frequency component is that of quasi-normal
ringing, {\it etc.}  For any such prior PDF, it 
is straightforward to calculate the corresponding maximum likelihood
estimator.  If the prior PDF has expected value ${\bf h}_0$ and
variance-covariance matrix ${\bf \Sigma}_0$, then the estimator is
\FL
\begin{equation}
{\bf h}_{\rm best-fit}({\bf s}) = \left[ {\bf \Sigma}^{-1} + {\bf
\Sigma}_0^{-1} \right]^{-1} \, \cdot \, \left[ {\bf \Sigma}^{-1} \cdot
{\bf s} + {\bf \Sigma}_0^{-1} \cdot {\bf h}_0 \right].
\label{bestfitsupervalue}
\end{equation}
Such a waveform estimator could be calculated numerically.

\subsection{Fidelity of waveform recovery}
\label{fidelity}

In this subsection we address the question of how close,
statistically, we expect our estimated waveform $h_{\rm best-fit}(t)$
to be to the original gravitational waveform $h(t)$. We can quantify
the closeness by means of the correlation coefficient
\begin{equation}
{\cal C} \equiv { \left( {\bf h} \, | \, {\bf h}_{\rm best-fit}
\right) \over
\sqrt{ \left( {\bf h} \, | \, {\bf h} \right)} \, 
\sqrt{ \left( {\bf h}_{\rm best-fit} \, | \, {\bf h}_{\rm best-fit}
\right)}},
\label{cdef}
\end{equation}
which takes values between $-1$ and $1$.  In appendix \ref{correl} we
show that for estimators of the form (\ref{bestfit}), the expected
value of ${\cal C}$ is approximately given by 
\begin{equation}
\langle {\cal C} \rangle \approx { \rho_{\rm bin} \over \sqrt{1 +
\rho_{\rm bin}^2}}, 
\label{cexpected}
\end{equation}
where $\rho_{\rm bin}^2$ is the matched filtering SNR squared per
frequency bin, given by 
\begin{equation}
\rho_{\rm bin}^2 = {\rho^2 \over {\cal N}_{\rm bins}}.
\end{equation}
Thus, as one would expect, the best-guess reconstructed waveform
agrees closely with the original gravitational waveform (${\cal C}$ is
close to $1$) when there is large SNR squared in each frequency bin,
and vice-versa \cite{notec}.   This result 
will also be approximately valid for waveforms obtained by simple
band-pass filtering when the duration $T$ of the signal satisfies $T
\gg 1/\Delta f$ (where $\Delta f$ is the frequency bandwidth of the signal).

Note that the quantity $\rho_{\rm bin}$ is to a good approximation just the
SNR which one obtains from band-pass filtering, from Eq.~(\ref{snrbp})
above.  Our criterion for the signal to be visible can therefore be
written as $\rho_{\rm bin} \agt 1$.  So our criteria for signal
visibility and for reconstructed signal fidelity turn out to be
essentially identical: the fidelity of signal reconstruction is good
when the merger signal is easily visible above the noise, as is fairly
obvious intuitively.

\section{USING INFORMATION PROVIDED BY REPRESENTATIVE
SUPERCOMPUTER SIMULATIONS}
\label{notmplsuper}

In this section we propose a computational strategy for numerical
relativists to pursue, if they successfully produce computer codes
capable of simulating BBH mergers, but if running such codes is too
expensive to permit an extensive survey of the merger parameter space.
In this case, for LIGO/VIRGO data analysis purposes, it would be
advantageous to do a very coarse survey of the parameter space aimed
at determining the ranges of several key parameters and at answering
several qualitative questions, as we now describe.

\begin{itemize}
\item
Do the waveforms contain a strong signature of an ``innermost stable
circular orbit'' (ISCO) \cite{kww,cook,laiwiseman}?  In the extreme
mass ratio limit $\mu \ll M$, there is such an orbit, and when the
smaller inspiralling black hole reaches it there is a transition from
a radiation-reaction-driven inspiral to a freely falling plunge
\cite{KipAmos}.  Correspondingly, there is a sharp drop in the
radiated energy per unit logarithmic frequency $d E / d (\ln f)$ at
the frequency corresponding to this orbit.  However, in the equal-mass
case, there may not be a sharp feature in the $d E / d (\ln f)$ plot,
if the timescale over which the orbital instability operates is
comparable to the radiation reaction timescale.  Or, if the spins of
the individual black holes are large and parallel to the orbital
angular momentum, the inspiral may smoothly join into the merger
without any plunge.  In the former case, the concept of ISCO would not
really be meaningful; and in the latter case, there would simply be
nothing resembling an ISCO in the evolution.  Simulations should be
able to settle this issue.

\item
A closely related question is: At what frequency does the adiabatic
approximation break down?  As seen in a coordinate system which
co-rotates with the black holes, the system evolves on a
radiation-reaction timescale which is initially much longer than the
orbital period \cite{Price,Pat}.  When does this separation of
timescales break down?  This separation of timescales underlies
proposed methods of calculating templates in the so-called
Intermediate Binary Black Hole (IBBH) regime after the post-Newtonian
approximation fails at $r \sim 12 M$ \cite{Price,Pat,Detweiler}.
Therefore, fully numerical templates will have to be used after the
adiabatic approximation fails.  Resolving this issue will probably
require exploration of both numerical relativity simulations and IBBH
calculations.  If the black holes' spins are small, one might expect
the transition point to coincide with estimates of the location of the
last stable circular orbit \cite{kww,cook,laiwiseman} around $r \sim 6
M$; our estimate (\ref{fmergedef}) of the frequency of the transition
from inspiral to merger roughly corresponds to this expectation.  But
with large spins, the system might evolve adiabatically all the way
into the merger.  (Note, however, that numerical relativity will still
be needed to model such evolution, whether it is adiabatic or not.)

\item
What is the approximate duration of the merger signal, and how does it
depend on the merger parameters such as the initial spins of the black
holes and the mass ratio?  The range of merger signal durations will
be an important input to algorithms for reconstructing the merger
waveform from the noisy data stream (see Sec.~\ref{bestguess}),
particularly in those cases in which the ringdown and/or inspiral
signals are too weak to be seen in the data stream.  Moreover, the
duration of the waveform (together with its bandwidth) approximately
determines the amount by which the SNR from band-pass filtering is
lower than the matched filtering SNR obtained with merger templates
[{\it cf}.\ Eq.\ (\ref{snrbp})].

\item
A closely related issue is: How much energy is radiated in the merger
waves relative to the ringdown waves?  Operationally, this question
reduces to asking what proportion of the total waveform produced
during the coalescence can be accurately fit by the ringdown's
decaying sinusoid.  In paper I we argued that if the spins of the
individual black holes are large and aligned with one another and with
the orbital angular momentum, 
then the system has too much angular momentum for it to be lost solely
through the ringdown, and that therefore the ringdown
waves should not dominate the merger.  On the other hand, if the spins of
the black holes are small, most of the radiated energy might well come
out in ringdown waves.

\item
What is the frequency bandwidth in which most of the merger waves'
power is concentrated?  In Ref.\ \cite{paperI} we assumed that when
one excises in the time domain the ringdown portion of the signal, the
remaining signal has no significant power at frequencies above the
quasi-normal ringing frequency of the final Kerr black hole.  However, this
assumption may not be valid.  As with the signal's duration, the range
of bandwidths of merger waveforms will be an input to algorithms for
reconstructing the merger waveform from the noisy data (see
Sec.~\ref{bestguess}), so this is an important issue.

\item
To what extent does the merger waveform chirp monotonically?  If we
represent the merger waves on a time-frequency wavelet basis, then we
know that at early times, the waves are concentrated at one frequency
with additional contributions in nearby harmonics.  At the end of the
merger signal, most of the power is concentrated near the frequency of
quasi-normal ringing of the final black hole.  One could extrapolate
in the time-frequency plane a line joining twice the orbital frequency
at the end of inspiral to the quasinormal ringing frequency at the
start of ringdown.  To what extent is the merger signal concentrated
near this line in the time-frequency plane?

\item
How much of the merger can be described as higher order QNR modes?  By
convention, we have been calling that phase of the coalescence which
is dominated by the most slowly damped, $l=m=2$ mode the ringdown
phase; but, before this mode dominates, QNR modes with different
values of $l$ and/or $m$ are likely to be present.  After the merger
has evolved to the point when the merged object can be accurately
described as a linear perturbation about a stationary black hole
background, there might or might not be any significant subsequent
period of time before the higher order modes have decayed away so much
as to be undetectable.  If simulations predict that higher order QNR
modes are strong for a significant period of time, then these higher
order QNR modes should be found by the normal ringdown search of the
data stream; no extra search should be needed.

\item
Does the merger signal have the property that we can distinguish a
``carrier waveform'' and a ``modulation''?  This separation would
require that the carrier waveform have a fairly large number of cycles
at a frequency well separated from that of the modulation.  It would
also require some mechanism to produce modulation, one possibility
being the precession of the black hole spins.  It is known that spin
precession does modulate the inspiral waveform {\cite{harisspin,kidder}},
and it is possible that a similar precession might be present during
at least part of the merger.

\end{itemize}

An improved understanding of these issues would be of use both
in extracting [{\it cf}.\ Sec.\ \ref{extendedmethod} above] and in
interpreting the merger waveforms.

\section{
INFORMATION OBTAINABLE FROM THE MERGER PHASE OF THE WAVES USING
TEMPLATES}
\label{infowithtemplates}

In the remainder of the paper we consider the optimistic scenario in
which a complete set of supercomputer generated theoretical merger
waveforms is available for data analysis.  In this section we describe in
qualitative terms the extra information that one can extract from the
merger waves using templates.  In Sec.\ \ref{tmplacc} we estimate how
accurate numerical templates need to be for data analysis purposes,
and in Sec.\ \ref{infosec} we estimate the total number of bits of
information obtainable from the merger waves using templates, and
discuss implications for the requirements on one's grid of templates.

If merger templates are available, it should be possible to perform
Wiener optimal filtering of the data stream for the merger signal,
just as will be done for the inspiral and ringdown signals.  When one
has no information about the BBH system, one would simply filter the
data with all numerical merger templates available, potentially a very
large number.  However, if the inspiral and/or the ringdown signals
have already been measured (as will be the case for most detected
signals), some information about the black hole 
binary's constituents will be available.   In such cases the total
number of merger 
templates needed will be reduced, perhaps substantially; one need
consider only those numerical templates whose parameters are
commensurate with the inspiral/ringdown measurements.

It may turn out that black hole mergers have such a wide variety of
behaviors that it will not be feasible to produce a complete family of
templates, even with a numerical code that can evolve mergers and
produce waveforms.  In such an eventuality, as mentioned in
Sec.~\ref{extractionsec} above, the
interpretation of an observed merger waveform could proceed as
follows:  The numerical relativists, with noisy data and numerical
code in hand, carry out a series of 
iterated numerical simulations, trying to produce a waveform that
matches the observed data.  (Clearly, it would be very useful for such
a procedure to have as much prior information as possible about the system's
parameters from the inspiral and/or ringdown phases, so
that the numerical relativists will know where in the binary black hole
parameter space to concentrate their computational efforts.)  Thus,
matched filtering might be possible even if the computation of a
complete set of template waveforms is too difficult to perform.

In attempting to match a merger template with gravitational-wave data,
one's primary goal would be to provide a test of general relativity
rather than the measurement of parameters.  A good match between the
measured waveform and a numerical template would constitute a strong
test of general relativity and provide the oft-quoted unambiguous
detection of black holes.  (Such an unambiguous detection could also
come from a measurement of the quasinormal ringing signal.)  Although
not the primary goal, matches between numerical merger templates and
the data stream would also be useful in measuring some of the system's
parameters, such as the total mass $M$ or the spin parameter $a$ of
the final black hole \cite{note3}.  These merger parameter
measurements could provide additional information about the source,
over and above that obtainable from the inspiral and ringdown signals.
For instance, in the second example discussed in Sec.\ \ref{priorinsp}
(a $30\,M_\odot$ BBH at $z=1$), the total redshifted mass $(1+z)M$
would be essentially unconstrained by the inspiral and ringdown
waveforms, but might be extractable from the measured merger waveform.
In other cases, a quantitative test of general relativity could be
obtained by verifying that parameters measured from the merger phase
are consistent with parameter measurements from the inspiral and
ringdown phases.

A close match between measured and predicted waveforms for BBH mergers
might also constrain some possible theories of gravity that generalize
general relativity.  Clifford Will has shown that the inspiral portion
of the waveform for neutron star-neutron star mergers will strongly
constrain the dimensionless parameter $\omega$ of Brans-Dicke theory
{\cite{cliff1}}.  Unfortunately, the most theoretically natural class of
generalizations of general relativity compatible with known
experiments, the so-called scalar-tensor theories {\cite{damour}}, may
not be strongly constrained (if at all) by measurements of BBH
mergers, since black holes, unlike neutron stars, cannot have any
scalar hair in such theories \cite{beckenstein}.

In order for the above endeavors to be successful, the numerical
templates must be sufficiently accurate.  In the next section, we turn
to a discussion of how accurate numerical templates need to be in
order to extract the information in merger signals.

\section{ACCURACY REQUIREMENTS FOR MERGER WAVEFORM TEMPLATES}
\label{tmplacc}

There will be unavoidable errors in the waveform templates produced by
supercomputer simulations, since these simulations are numerical.
Suppose that the physical waveform for some particular source is
$h(t;{\bftheta})$, where the components of the vector ${\bftheta} =
(\theta^1, \ldots, \theta^{n_p})$ represent the various parameters
upon which the waveform depends.  Then, a simulation of the evolution
of that source will predict a slightly different waveform
$h(t;{\bftheta}) + \delta h(t;{\bftheta})$, where $\delta
h(t;{\bftheta})$ is the numerical error.  One would like the numerical
error to be small enough not to have a significant effect on signal
searches, parameter extraction or any other types of data analysis
that might be carried out using the template waveforms.  In this
section we suggest an approximate rule of thumb [Eqs.~(\ref{Deltadef})
and (\ref{accuracycriterion})] for estimating when numerical
errors are sufficiently small, and discuss its meaning and derivation.

\subsection{Accuracy criterion and implementation}
\label{accimpl}

The accuracy criterion can be simply expressed in terms of the inner
product introduced in Sec.\ \ref{notations} above [which is defined by
Eq.~(\ref{innerprod}) or alternatively by Eqs.\
(\ref{Gammadef})---(\ref{discreteinnerprod})]: For a given template
$h(t)$, our rule of thumb is that the numerical error $\delta h(t)$
should be small enough that the quantity
\begin{equation}
\Delta \equiv {1 \over 2}  { \left( \delta h | \delta h \right)
\over \left( h | h \right) } 
\label{Deltadef}
\end{equation}
satisfies
\begin{equation}
\Delta \alt 0.01.
\label{accuracycriterion}
\end{equation}
[The fractional loss in event detection rate in signal searches is
$\sim 3 \Delta$, so the value of $0.01$ in
Eq.~(\ref{accuracycriterion}) is chosen to correspond to a 3\% loss in
event rate; see Sec.~\ref{derivationacc} below].
For the purpose of evaluating the inner product numerically, note that
the absolute normalization of the noise 
spectrum $S_h(f)$ is unimportant, and that one could use, for example,
Eqs.~(4.1)---(4.3) of Ref.~\cite{paperI} to specify the shape of the
noise spectrum.

In practice, Eq.~(\ref{accuracycriterion}) translates to a
fractional accuracy per data point $h_j = h(t_j)$ of about $0.01 /
\sqrt{{\cal N}_{\rm points}}$, where ${\cal N}_{\rm points}$ is
the number of numerical data points used to describe the templates, if
the errors at each data point are effectively uncorrelated.  If,
however, these errors add coherently in the integral (\ref{Deltadef}),
the requirement on fractional accuracy at each data point will be more
stringent.

It should be straightforward in principle to ensure that numerical
templates satisfy the criterion (\ref{accuracycriterion}).  Let us
schematically denote a numerically generated template as $h_{\rm
num}(t,\varepsilon)$, where $\varepsilon$ represents the set of
tolerances (grid size, size of time steps, {\it etc}.) that govern the
accuracy of the numerical calculation.  (Representing this set of
parameters by a single parameter $\varepsilon$ is an
oversimplification but is adequate for the purposes of our
discussion.)  One can then iterate one's calculations varying the
parameter $\varepsilon$ in order to obtain templates that are
sufficiently accurate, using the following standard type of procedure:
First, calculate the template $h_{\rm num}(t,\varepsilon)$.  Second,
calculate the more accurate template $h_{\rm
num}(t,\varepsilon^\prime)$ for some choice of $\varepsilon^\prime <
\varepsilon$, for example $\varepsilon^\prime = \varepsilon/2$.
Third, make the identifications
\begin{eqnarray}
h(t) &\equiv& h_{\rm num}(t,\varepsilon^\prime),\nonumber\\
\delta h(t) &\equiv & h_{\rm num}(t,\varepsilon^\prime)
- h_{\rm num}(t,\varepsilon)
\end{eqnarray}
and insert these quantities into Eq.~(\ref{Deltadef}) to calculate
$\Delta$.  This allows one to assess the accuracy of the template
$h_{\rm num}(t,\varepsilon)$.  Finally, iterate this procedure until
Eq.\ (\ref{accuracycriterion}) is satisfied.

\subsection{Derivation and meaning of accuracy criterion}
\label{derivationacc}

The required accuracy for the numerical templates depends on how and
for what purpose those templates are used.  As discussed in the
Introduction, merger templates might be used in several different
ways:

\begin{itemize}

\item
They might be used as search templates for signal searches using
matched filtering.  Such searches will probably not be feasible, at
least initially, as they would require the computation of an
inordinately large number of templates.  

\item
For BBH events that have already been detected via matched filtering
of the inspiral or ringdown waves, or by the noise-monitoring
detection technique \cite{paperI,ComingSoonToAJournalNearYou} applied
to the merger waves, the merger templates might be used for matched
filtering in order to measure the binary's parameters and test general
relativity.  This use of merger templates could correspond to the
third scenario that was discussed in Sec.\ \ref{extractionsec}, where
iterated runs of the supercomputer codes are performed to produce a
template that best fits a dataset known to contain BBH merger
gravitational waves.  This scenario would not require that a complete
set of templates be computed and stored, and thus is somewhat more
feasible than matched filtering signal searches using the merger
waves.

\item If one has only a few, representative supercomputer simulations
and their associated waveform templates at one's disposal,
one might simply perform a qualitative comparison between the measured
waveform and templates in order to deduce qualitative information
about the BBH source.  This is the second scenario described in Sec.\
\ref{extractionsec}.

\end{itemize}

In this section we estimate the accuracy requirements for the first
two of these uses of merger templates.  

Consider first signal searches using matched filtering.  The expected
SNR $\rho$ obtained for a gravitational waveform $h(t)$ when using a
template waveform $h_T(t)$ is given by
\cite{owen} 
\begin{equation}
\rho = {\left( h | h_T \right) \over \sqrt { \left( h_T | h_T \right)
}}.
\label{rhoacc}
\end{equation}
If we substitute $h_T(t) = h(t) + \delta h(t)$ into Eq.~(\ref{rhoacc})
and expand to second order in $\delta h$, we find that the fractional
loss $\delta \rho / \rho$ in SNR produced by the numerical
error $\delta h(t)$ is given by
\begin{equation}
{\delta \rho \over \rho} = \Delta_1 + O[(\delta h)^3],
\label{ans1}
\end{equation}
where
\begin{equation} 
\Delta_1 \equiv {1 \over 2} \left[ { \left( \delta h | \delta h \right)
\over \left( h | h \right) } - {\left( \delta h | h \right)^2 \over
\left( h|h \right)^2 } \right].
\label{Delta1def}
\end{equation}
Note that the quantity $\Delta_1$ is proportional to $(\delta h_1 | 
\delta h_1)$, where $\delta h_1$ is the component of 
$\delta h$ perpendicular to $h$ with respect to the inner product
(\ref{discreteinnerprod}).  Thus, a numerical error of the form
$\delta h(t) \propto h(t)$ will not contribute to the fractional loss
(\ref{ans1}) in SNR.  This is to be expected, since
the quantity (\ref{rhoacc}) is independent of the absolute
normalization of the templates $h_T(t)$.

Now, the event detection rate is proportional to the cube of the
SNR, and hence the fractional loss in event
detection rate that results from using inaccurate numerical templates
is approximately $3\delta \rho/\rho$ \cite{owen}.  If one demands that
the fractional loss in event rate be less than, say, $3 \%$, then one
obtains the criterion \cite{noteowen}
\begin{equation}
\Delta_1 \le 0.01.
\label{accuracycriterion1}
\end{equation}
It is clear from Eqs.~(\ref{Deltadef}) and (\ref{Delta1def}) that
$\Delta_1 \le \Delta$.  Hence, the condition
(\ref{accuracycriterion1}) is less stringent than the condition
(\ref{accuracycriterion}) above.  The justification for imposing the
more stringent criterion (\ref{accuracycriterion}) rather than
(\ref{accuracycriterion1}) derives from the use of templates for
parameter extraction.

Consider next using merger templates for the purpose of measuring
parameters via matched filtering.  In principle, one could hope to
measure all of the 15 parameters on which the merger waveforms depend
by combining the outputs of several detectors with a complete bank of
templates (although in practice the accuracy with which some of those
15 parameters can be measured is not likely to be very good).  In the
next few paragraphs we derive an approximate condition on $\Delta$
[Eq.~(\ref{acc3})] which results from demanding that the systematic
errors in the measured values of all the parameters be small compared
to the statistical errors due to detector noise.  (We note that one
would also like to use matched filtering to test general relativity
with these waves; the accuracy criterion that we derive for parameter
measurement will also approximately apply to tests of general
relativity.)

Let the gravitational waveform be $h(t; \,\bftheta)$, where $\bftheta =
(\theta^1, \ldots, \theta^{n_p})$.  Let ${\hat \theta}^\alpha$, $1 \le \alpha
\le n_p$, be the best-fit values of $\theta^\alpha$ 
given by the matched-filtering process.  The quantities ${\hat
\theta}^\alpha$ depend on the detector noise and are thus random
variables. In the high SNR limit, the variables ${\hat\theta}^\alpha$
have a multivariate Gaussian distribution with (see, {\it e.g.},
Ref.~\cite{cutlerflan94})
\begin{equation}
\langle \delta {\hat \theta}^\alpha \,\,\delta {\hat \theta}^\beta \rangle =
\Sigma^{\alpha\beta},
\label{staterror}
\end{equation}
where $\delta {\hat \theta}^\alpha \equiv {\hat \theta}^\alpha -
\langle {\hat \theta}^\alpha \rangle$ and the matrix
$\Sigma^{\alpha\beta}$ is defined after Eq.~(\ref{metricTH}).  The
systematic error $\Delta \theta^\alpha$ in the inferred values of the
parameters $\theta^\alpha$ due to the template error $\delta h$ can be
shown to be approximately
\begin{equation}
\Delta \theta^\alpha = \Sigma^{\alpha\beta} \, \left( {\partial h
\over \partial 
\theta^\beta} \bigg| \delta h \right).
\label{systerror}
\end{equation}
From Eqs.~(\ref{staterror}) and (\ref{systerror}) we find that in
order to guarantee that the systematic error in each of the 
parameters be smaller than some number $\varepsilon$ times that
parameter's statistical error, we must have
\begin{equation}
||\delta h_\parallel ||^2 \equiv \left( \delta h_\parallel | \delta
h_\parallel \right) \le \varepsilon^2. 
\label{acc1}
\end{equation}
Here $\delta h_\parallel$ is the component of $\delta h$ parallel to
the tangent space of the manifold of signals ${\cal S}$
$h(t,\bftheta)$ discussed in Sec.~\ref{notations}.  It is given by 
\begin{equation}
\delta h_\parallel = \Sigma^{\alpha\beta} \, \left( \delta h \bigg|
{\partial h \over \partial \theta^\alpha} \right) \, {\partial h \over
\partial \theta^\beta}. 
\end{equation}

The magnitude $|| \delta h_\parallel ||$ of this component of $\delta
h$ depends on details of the number of parameters, and on how the
waveform $h(t,{\bftheta})$ varies with these parameters.  However, a
strict upper bound is given by
\begin{equation}
|| \delta h_\parallel || \le \, ||\delta h||.
\label{acc2}
\end{equation}
If we combine Eqs.~(\ref{Deltadef}), (\ref{acc1}), and (\ref{acc2}) we
obtain the condition
\begin{equation}
\Delta \le {\varepsilon^2 \over 2 \rho^2}.
\label{acc3}
\end{equation}
If we insert reasonable estimates for $\rho$ and
$\varepsilon$---namely, $\rho
\simeq 7$, $\varepsilon \simeq 1$---we recover the criterion
(\ref{accuracycriterion}).  [Note that the requirement (\ref{acc3}) is
probably rather more stringent than need be: the left hand side of
Eq.~(\ref{acc2}) is likely smaller than the right hand side by a
factor $\sim \sqrt{n_p/{\cal N}_{\rm bins}}$, where $n_p$ is the
number of parameters and ${\cal N}_{\rm bins}$ is the dimension of the
total space of signals $V$.]

In Sec.~\ref{infosec} below we give an alternative derivation of the
accuracy criterion (\ref{acc3}) using information theory.

The expected order of magnitude $\rho \simeq 7$ of the SNR that leads
to the criterion (\ref{accuracycriterion}) is appropriate for ground
based interferometers such as LIGO and VIRGO \cite{paperI}.  However,
for the space-based LISA interferometer, much higher SNRs are
expected; see, {\it e.g.}, Ref.~\cite{paperI}.  Correspondingly,
numerical templates used for testing relativity and measuring
parameters with LISA data will have to be substantially more accurate
than those used with data from ground based instruments.

\section{
NUMBER OF BITS OF INFORMATION OBTAINABLE FROM THE MERGER SIGNAL AND
IMPLICATIONS FOR TEMPLATE CONSTRUCTION}
\label{infosec}

In this section, we describe how to use information theory to quantify
how much can be learned from a gravitational-wave measurement.  In
information theory, a quantity called ``information'' (analogous to
entropy) can be associated with any measurement process: it is simply
the base 2 logarithm of the number of distinguishable outcomes of the
measurement \cite{brillouin,coverthomas}.  Equivalently, it is the
number of bits required to store the knowledge gained from the
measurement.  Here we specialize the notions of information theory to
gravitational wave measurements, and estimate the number of bits of
information which one can gain in different cases.

Let us first consider the situation in which templates are
unavailable.  Suppose that our prior information describing the signal
is that it lies inside some frequency band of width $\Delta f$ say,
that it lies inside some time interval of length $T$ say.
We will denote by $I_{\rm total}$ the base 2
logarithm of the number of waveforms ${\bf h}$ that are
distinguishable by the measurement, that are compatible with our prior
information, and that are compatible with our measurement of the
detector output's magnitude $\rho({\bf s}) = ||{\bf s}||$
\cite{noteinfo}.  We give a precise version of this definition in
Sec.~\ref{Itotalsec} below [Eq.~(\ref{infodef})].  Note that the vast
majority of these $2^{I_{\rm total}}$ waveforms are completely
irrelevant to BBH mergers; the BBH merger signals are a small subset
(the manifold ${\cal S}$) of all distinguishable waveforms with the
above characteristics.  However, without prior information about which
waveforms are relevant, we cannot {\it a priori} ignore any waveform,
and so we must include in our counting even the irrelevant ones.  Note
also that the quantity $I_{\rm total}$ quantifies the amount of
information we gain from the measurement about the shape of the merger
waveform; however, in the absence of any templates we do not learn
anything about the source of waves.  In Appendix \ref{app_info} we
derive and in Sec.\ \ref{Itotalsec} we discuss an approximate formula
for $I_{\rm total}$ in terms of the matched filtering signal-to-noise
ratio $\rho$ and the number of frequency bins ${\cal N}_{\rm bins}$
[Eq.~(\ref{infoapprox})].  This approximate formula can be understood
with a simple, intuitive argument, which we also elucidate in Sec.\
\ref{Itotalsec}.

Consider now the situation in which templates are available.  In
Sec.~\ref{Isourcesec} below [Eq.~(\ref{infodef1})] we define a
quantity $I_{\rm source}$, 
which is, roughly speaking, the base 2 logarithm of the number of
distinguishable waveforms that could have come from BBH mergers and
that are distinguishable in the detector noise.  The quantity $I_{\rm
source}$ differs from the quantity $I_{\rm total}$ in that it
counts only the subset of waveforms relevant to BBH mergers.
Note that the information which $I_{\rm source}$ quantifies is
information about the {\it source} of the waves: when templates are
available we can relate the waveform shape to properties of the BBH
system.  In Appendix \ref{app_info} we derive an approximate formula
[Eq.~(\ref{infoapprox1})] for $I_{\rm source}$. 

Finally, in Sec.~\ref{infolosssec} we estimate how much of the
information $I_{\rm source}$ is lost due to template numerical error
[Eq.~(\ref{anss})] 
and due to having insufficiently many templates in one's grid
[Eq.~(\ref{anss3})], and
deduce requirements one's grid of templates must satisfy in order for
the loss of information to be unimportant.

\subsection{Total information gain}
\label{Itotalsec}

A precise definition of the total information gain $I_{\rm
total}$ is the following:  
Let $T$ and $\Delta f$ be a priori upper bounds for the durations and
bandwidths of merger signals, and let $V$ be the vector space of
signals with duration $\le T$ inside the relevant frequency band.
This vector space $V$ has dimension ${\cal N}_{\rm bins} = 2 T \Delta
f$.  Let ${\bf s}$,
${\bf h}$ and ${\bf n}$ denote the detector output, gravitational wave
signal and detector noise respectively, so that ${\bf s} = {\bf h} +
{\bf n}$.  The quantities ${\bf s}$, ${\bf h}$, and ${\bf n}$ are all
elements of $V$.  Let $p^{(0)}({\bf h})$ be the PDF describing our
prior information about the gravitational wave signal \cite{noteprior}, 
and let $p({\bf h}\,|\,{\bf s})$ denote the posterior PDF for ${\bf
h}$ after the 
measurement, {\it i.e.}, the PDF for ${\bf h}$ given that the detector
output is ${\bf s}$.  A standard Bayesian analysis
shows that $p({\bf h}\,|\,{\bf s})$ will be
given by 
\begin{equation}
p({\bf h}\,|\,{\bf s}) = {\cal K} \, p^{(0)}({\bf h}) \, \exp\left[
-\left( {\bf s} - {\bf h} \,|\, {\bf s} - {\bf h} \right) /2 \right]
\label{standardb}
\end{equation}
where ${\cal K}$ is a normalization constant \cite{finnmeasure}.
Finally, let $p[{\bf h} \,|\, \rho({\bf s})]$ be the PDF of ${\bf h}$
given that the magnitude $||{\bf s}||$ of the measured signal is
$\rho({\bf s})$.  We define the quantity $I_{\rm total}$ to be
\begin{equation}
I_{\rm total} \equiv \int d{\bf h} \,\, p({\bf h} \,|\, {\bf s})
\log_2 \left[ { p({\bf h} 
\,|\, {\bf s}) \over p[{\bf h} \,|\, \rho({\bf s})]} \right].
\label{infodef}
\end{equation}
By this definition, $I_{\rm total}$ is the {\it relative information}
of the probability distributions $p[{\bf h}\,|\,\rho({\bf s})]$ and $p({\bf
h}\,|\,{\bf s})$ \cite{coverthomas}.  
In Appendix \ref{app_info} we show
that the quantity (\ref{infodef}) in fact represents the base 2
logarithm of the number of distinguishable wave shapes 
that could have
been measured and that are compatible with one's measurement of the
magnitude $\rho({\bf s})$ of the data stream \cite{noteinfo}.  Thus,
one learns  $I_{\rm total}$ bits of information 
about the waveform ${\bf h}$ when one goes from knowing only the
magnitude $||{\bf 
s}||$ of the detector output to knowing the actual detector output
${\bf s}$.

We also show in Appendix \ref{app_info} that in the
limit of no prior information other than $T$ and
$\Delta f$, an approximate formula for the quantity
(\ref{infodef}) is
\begin{equation}
I_{\rm total} = {1 \over 2} {\cal N}_{\rm bins} \, \log_2 \left[
\rho({\bf s})^2 / {\cal N}_{\rm bins} \right] + O[\ln {\cal N}_{\rm
bins}].
\label{infoapprox0}
\end{equation} 
The formula (\ref{infoapprox0}) is valid in the limit of large ${\cal
N}_{\rm bins}$ for fixed $\rho({\bf s})^2/{\cal N}_{\rm bins}$, and
moreover applies only when
\begin{equation}
\rho({\bf s})^2 / {\cal N}_{\rm bins} > 1;
\label{condtvalidity}
\end{equation}
see below for further discussion of this point.

There is a simple and intuitive way to understand the result
(\ref{infoapprox0}).  Let us fix the gravitational waveform, ${\bf
h}$, considered as a point in the ${\cal N}_{\rm bins}$-dimensional
Euclidean space $V$.  What is measured is the detector output ${\bf h}
+ {\bf n}$, whose location in $V$ is displaced from that of ${\bf h}$.  The
direction and magnitude of the displacement depend upon the particular
instance of the noise ${\bf n}$.  However, if we average over an ensemble
of realizations of the noise, we can see that the displacement due to
the noise is in a 
random direction and has rms magnitude $\sqrt{{\cal N}_{\rm bins}}$
(since on an appropriate basis each component of ${\bf n}$ has rms
value 1).  Therefore, all points ${\bf h}^\prime$ lying inside a
hypersphere of radius $\sqrt{{\cal N}_{\rm bins}}$ centered on ${\bf
h}$ are effectively indistinguishable from each other.  The volume of
such a hypersphere is
\begin{equation}
C_{{\cal N}_{\rm bins}} (\sqrt{{\cal N}_{\rm bins}})^{{\cal N}_{\rm bins}},
\label{vol1}
\end{equation}
where $C_{{\cal N}_{\rm bins}}$ is a constant whose value is
unimportant.  When we measure a detector output ${\bf s}$ with
magnitude $\rho({\bf s})$, the set of signals ${\bf h}$ that could
have given rise to an identical measured $\rho({\bf s})$ will form a
hypersphere of radius $\sim \rho({\bf s})$ and volume
\begin{equation}
C_{{\cal N}_{\rm bins}} \rho({\bf s})^{{\cal N}_{\rm bins}}.
\label{vol2}
\end{equation}
The number of distinguishable signals in this large hypersphere will
be approximately the ratio of the two volumes (\ref{vol1}) and
(\ref{vol2}); the base 2 logarithm of this ratio is the quantity
(\ref{infoapprox0}).

Equation (\ref{infoapprox0}) expresses the information gain as a
function of the magnitude of the measured detector output ${\bf s}$.
We now re-express this information gain in terms of properties of the
gravitational wave signal ${\bf h}$.  For a given ${\bf h}$, Eqs.\
(\ref{magnitude1}) and (\ref{magnitude2}) show that the detector
output's magnitude $\rho({\bf s})$ will be approximately
\begin{equation}
\rho({\bf s})^2 \approx \rho^2 + {\cal N}_{\rm bins} \pm \sqrt{{\cal N}_{\rm
bins}}.
\label{magnitude3}
\end{equation}
Here $\rho^2 = ||{\bf h}||^2$ is the SNR squared (\ref{snr}) that
would be achieved if matched filtering were possible (if templates
were available).  We use $\rho$ simply as a convenient measure of
signal strength; in this context, it is meaningful even in situations
where templates are unavailable and where matched filtering cannot be
carried out.  The last term in Eq.~(\ref{magnitude3}) gives the
approximate size of the statistical fluctuations in $\rho({\bf s})^2$.
We now substitute Eq.~(\ref{magnitude3}) into
Eq.~(\ref{infoapprox0}) and obtain
\begin{eqnarray}
I_{\rm total}  &=& {1 \over 2} {\cal N}_{\rm bins} \, \log_2 \left[
1 + \rho^2 / {\cal N}_{\rm bins} \right] \nonumber \\
\mbox{} && \times  \left[ 1 + O\left( {\ln {\cal
N}_{\rm bins} \over {\cal N}_{\rm bins}}\right) + O\left( {1 \over
\sqrt{{\cal N}_{\rm bins}}} \right) \right].
\label{infoapprox}
\end{eqnarray} 
Also, the condition (\ref{condtvalidity}) for the applicability of
Eq.~(\ref{infoapprox0}), when expressed in terms of $\rho$ instead of
$\rho({\bf s})$, becomes 
\begin{equation}
{\rho^2 \over {\cal N}_{\rm bins}} \pm {1 \over \sqrt{{\cal N}_{\rm
bins}}} \ge 0,
\end{equation}
which will be satisfied with high probability when $\rho \gg {\cal
N}_{\rm bins}^{1/4}$ \cite{notedetect1}.  In the regime $\rho \alt
{\cal N}_{\rm bins}^{1/4}$, the condition (\ref{condtvalidity}) is
typically not satisfied and the formula (\ref{infoapprox0}) does not
apply; we show in Appendix \ref{app_info} that in this case the
information gain (\ref{infodef}) is usually very small, depending
somewhat on the prior PDF $p^{(0)}({\bf h})$.  [In contexts other than
BBH merger waveforms, the information gain can be large in the regime
$\rho \ll {\cal N}_{\rm bins}^{1/4}$ if the prior PDF $p^{(0)}({\bf
h})$ is very sharply peaked.  For example, when one considers
measurements of binary neutron star inspirals with advanced LIGO
interferometers, the  
information gain in the measurement is large even though typically one
will have $\rho \ll {\cal N}_{\rm bins}^{1/4}$, because we have very
good prior information about inspiral waveforms.]

As an example, a typical detected BBH event might have an SNR for the
merger signal of $\rho \sim 10$, and the number of frequency bins
${\cal N}_{\rm bins}$ might be $\sim 30$ \cite{paperI}.  Then,
Eq.~(\ref{infoapprox}) tells us that $\sim 3 \times 10^9 \approx
2^{32}$ signals of the same magnitude could have been distinguished,
thus the number of bits of information gained is $\sim 32$.  More
generally, for ground based interferometers we expect $\rho$ to lie in
the range $5 \alt \rho \alt 100$ \cite{paperI}, and therefore we
expect $10 \alt I_{\rm total} \alt 120$; and for LISA we expect $\rho$
to typically lie in the range $10^3 \alt \rho \alt 10^5$ so that $200
\alt I_{\rm total} \alt 400$.

\subsection{Amount of information gained about the wave's source}
\label{Isourcesec}

Consider now the idealized situation in which a complete family of
accurate theoretical template waveforms ${\bf h}(\bftheta)$ are
available for the merger.  Without templates, we gain $I_{\rm total}$
bits of information about the shape of the gravitational waveform in a
measurement.  With templates, some---but not all---of this information
can be translated into information about the BBH source.  For
instance, suppose in the example considered above that the number of
distinguishable waveforms that could have come from BBH mergers and
that are distinguishable in the detector noise is $2^{25}$.  
(This number must be less that the total number $\sim 2^{32}$ of
distinguishable waveform shapes, since waveforms from BBH mergers will
clearly not fill out the entire function space $V$ of possible
gravitational waveforms.)  In this example, by
identifying which template best fits the detector output, we can gain
$\sim 25$ bits of information about the BBH source (e.g.~about the
black holes' masses or spins).  We will call this number of bits of
information $I_{\rm source}$; clearly $I_{\rm source} \le I_{\rm
total}$ always.

What of the remaining $I_{\rm total} - I_{\rm source}$ bits of
information (7 bits in the above example)?  If the detector output is
close to one of the template shapes, then this closeness can be
regarded as evidence in favor of the theory of gravity (general
relativity) used to compute the templates, so the $I_{\rm total} -
I_{\rm source}$ extra bits of information can be viewed as information
about the validity of general relativity.  If one computed templates
in more general theories of gravity, one could in principle translate
these $I_{\rm total} - I_{\rm source}$ bits of information into a
quantitative form and obtain 
constraints on the parameters entering into the gravitational theory.
However, with only general-relativistic templates at one's disposal,
the information contained in the $I_{\rm total} - I_{\rm source}$ bits
will simply result in a qualitative confirmation of general
relativity, in the sense that one of the general relativistic
templates will provide a good fit to the data.

It is possible to give a precise definition of the number of bits of
information gained about the BBH source, $I_{\rm source}$, in the
following way.  Let $p(\bftheta \, | \, {\bf s})$ denote the
probability distribution for the source parameters $\bftheta$ given
the measurement ${\bf s}$.  This PDF is given by a formula analogous
to Eq.~(\ref{standardb})
\cite{finnmeasure} 
\begin{equation}
p(\bftheta \,|\,{\bf s}) =  {\cal K} \, p^{(0)}(\bftheta) \, \exp\left[
-\left( {\bf s} - {\bf h}(\bftheta) \,|\, {\bf s} - {\bf h}(\bftheta)
\right) /2 \right], 
\label{standardb1}
\end{equation}
where $p^{(0)}(\bftheta)$ is the prior PDF for $\bftheta$ and ${\cal
K}$ is a normalization constant.  Let $p[\bftheta \, | \, \rho({\bf
s})]$ be the posterior PDF for $\bftheta$ given that the magnitude
$||{\bf s}||$ of the measured signal is $\rho({\bf s})$.  Then we
define
\begin{equation}
I_{\rm source} \equiv \int d{\bftheta} \,\, p({\bftheta} \,|\, {\bf s})
\log_2 \left[ { p({\bftheta} 
\,|\, {\bf s}) \over p[{\bftheta} \,|\, \rho({\bf s})]} \right].
\label{infodef1}
\end{equation}

The number of bits of information (\ref{infodef1}) gained about the
BBH source will clearly depend on the details of how the gravitational
waveforms depend on the source parameters, on the prior expected
ranges of these parameters, {\it etc.}  In Appendix \ref{app_info} we
argue that to a rather crude approximation, $I_{\rm source}$ should be
given by the formula (\ref{infoapprox}) with ${\cal N}_{\rm bins}$
replaced by the number of parameters ${\cal N}_{\rm param}$ on which
the waveform has a significant dependence:
\begin{equation}
I_{\rm source} \approx {1 \over 2} {\cal N}_{\rm param} \, \log_2
\left[ 1 + \rho^2 / {\cal N}_{\rm param} \right].
\label{infoapprox1}
\end{equation} 
Note that the quantity ${\cal N}_{\rm param}$ should be bounded above
by the quantity $n_p$ discussed in Sec.~\ref{notations}, but may be
somewhat smaller than $n_p$.  This will be the case if the waveform
depends only very weakly on some of the parameters $\theta^\alpha$.
Equation (\ref{infoapprox1}) is only valid when ${\cal N}_{\rm param}
\le {\cal N}_{\rm bins}$.  For BBH mergers we expect ${\cal N}_{\rm
param} \alt 15$, which from Eq.~(\ref{infoapprox1}) predicts that
$I_{\rm source}$ lies in the range $\sim 10$ bits to $\sim 70$ bits
for signal-to-noise ratios $\rho$ in the range $5$ to $100$ (the
expected range for ground based interferometers \cite{paperI}), and
$\sim 100$ bits to $\sim 200$ bits for $\rho$ in the range $10^3$ to
$10^5$ expected for LISA \cite{paperI}.

\subsection{Loss of information about source due to template
inaccuracies or to sparseness of the lattice of templates}
\label{infolosssec}

As we discussed in Sec.\ \ref{tmplacc}, numerical templates will
contain some unavoidable error due to the calculational technique.  In
this section we analyze how that error affects the information gained
in the measurement process, and use this analysis to infer the maximum
allowable template error.

Let us write
\begin{equation}
{\bf h}_T(\bftheta) = {\bf h}(\bftheta) + {\bf \delta h}(\bftheta),
\end{equation}
where ${\bf h}(\bftheta)$ denotes the true waveform shape, ${\bf
h}_T(\bftheta)$ the numerical template, and ${\bf \delta h}(\bftheta)$
the numerical error.  It is clear that the numerical error will reduce
the amount of information (\ref{infodef1}) one can obtain about the
source.  We can make a crude estimate of the amount of reduction in
the following way.  We model the numerical error as a random process
with
\begin{equation}
\langle \delta h_i \, \delta h_j \rangle = {\cal C}_{ij},
\end{equation}
where for simplicity we take ${\cal C}_{ij} = \lambda \Gamma_{ij}$ for
some constant $\lambda$.  Here $\Gamma_{ij}$ is the matrix introduced
in Eq.\ (\ref{Gammadef}).  The expected value of $\left( {\bf \delta
h} \, | \, {\bf \delta h} \right)$ is then given by, from Eq.\
(\ref{discreteinnerprod}),
\begin{eqnarray}
\langle \, \left( {\bf \delta h} \, | \, {\bf \delta h} \right) \,
\rangle &=& \Sigma^{ij} \langle \delta h_i \delta h_j \rangle
\nonumber \\
\mbox{} &=& \Sigma^{ij} \, \lambda \Gamma_{ij} = \lambda {\cal N}_{\rm bins},
\label{duvet}
\end{eqnarray}
where we have used Eq.~(\ref{Sigmadef}).
We can write $\lambda$ in terms of the quantity $\Delta$ discussed in
Sec.~\ref{tmplacc} by combining Eqs.~(\ref{Deltadef}) and
(\ref{duvet}), yielding
\begin{equation}
\lambda = 2 \Delta { \rho^2 \over {\cal N}_{\rm bins}}.
\label{lambdadef}
\end{equation}

The information $I_{\rm source}^\prime$ which one obtains when
measuring with inaccurate templates can be calculated by treating the
sum of the detector noise ${\bf n}$ and the template numerical error
${\bf \delta h}$ as an effective noise ${\bf n}^{({\rm eff})}$.  This
effective noise is characterized by the covariance matrix
\begin{equation}
\langle n_i^{({\rm eff})} \, n_j^{({\rm eff})} \rangle = \Gamma_{ij} +
\lambda \Gamma_{ij}.
\end{equation}
Thus, in this simplified model, the effect of the numerical error is
to increase the noise by a factor $1 + \lambda$.  The new
information gain $I_{\rm source}^\prime$ is therefore given by
Eq.~(\ref{infoapprox1}) with $\rho$ replaced by an effective SNR
$\rho^\prime$, where 
\begin{equation}
(\rho^\prime)^2 = { \rho^2 \over 1 + \lambda }.
\label{rhoprimedef}
\end{equation}
If we now combine Eqs.~(\ref{infoapprox1}), (\ref{lambdadef}) and
(\ref{rhoprimedef}), we find that the loss in information due to
template inaccuracy
\begin{equation}
\delta I_{\rm source} = I_{\rm source} - I_{\rm source}^\prime
\end{equation}
is given by
\FL
\begin{equation}
\delta I_{\rm source} = \rho^2 \left({\rho^2 \over {\cal N}_{\rm
param} + \rho^2} \right) \, \left( {{\cal N}_{\rm param} \over
{\cal N}_{\rm bins}} \right) \Delta + O(\Delta^2). 
\label{anss}
\end{equation}
To ensure that $\delta I_{\rm source} \alt 1$ bit, we therefore must
have
\begin{equation}
\Delta \alt {1 \over \rho^2} \, \left( { {\cal N}_{\rm param} +
\rho^2 \over \rho^2} \right) \, \left( {{\cal N}_{\rm bins} \over
{\cal N}_{\rm param} }\right).
\label{anss1}
\end{equation}
This condition is a more accurate version of the condition
(\ref{acc3}) that was derived in Sec.~\ref{tmplacc}.  It approximately
reduces to the condition (\ref{acc3}) for typical BBH events
(except in the unrealistic limit $\rho^2 \ll {\cal N}_{\rm
param}$), since ${\cal N}_{\rm param} \sim 10$ 
and $10 \alt {\cal N}_{\rm bins} \alt 100$ \cite{paperI}.

Turn next to the issue of the required degree of fineness of a
template lattice; {\it i.e.}, the issue of how close in parameter space
successive templates must be 
to one another.  This is mostly relevant to the third scenario
described in Sec.\ \ref{extractionsec}, in which numerical relativists
are able to simulate essentially arbitrary BBH mergers, and to carry
out a large number of such simulations. 
We can parameterize the degree of fineness by a dimensionless
parameter $\varepsilon_{\rm grid}$ in the following way: the lattice
is required to have the property that for any possible true signal
${\bf h}(\bftheta)$, there exists some template ${\bf h}(\bftheta^*)$
in the lattice with
\begin{equation}
{\left( {\bf h}(\bftheta) \, | \, {\bf h}(\bftheta^*) \right) 
\over 
\sqrt{ \left( {\bf h}(\bftheta) \, | \, {\bf h}(\bftheta) \right) } \,\,
\sqrt{\left( {\bf h}(\bftheta^*) \, | \, {\bf h}(\bftheta^*) \right)}} \,
\ge 1 - \varepsilon_{\rm grid}.
\label{griddef}
\end{equation}
The quantity $1 - \varepsilon_{\rm grid}$ is called the minimal match
\cite{owen}.  Suppose that one defines a metric on the space $V$ of
templates using the norm (\ref{norm}).  It then follows from
Eq.~(\ref{griddef}) that the largest possible distance $D_{\rm max}$
between an incoming signal ${\bf h}(\bftheta)$ and some rescaled
template ${\cal A} {\bf h}(\bftheta^*)$ with ${\cal A}>0$ is
\begin{equation}
D_{\rm max} = \sqrt{2 \varepsilon_{\rm grid}} \, \rho,
\end{equation}
where $\rho$ is the matched filtering SNR (\ref{snr}) of the incoming
signal.

We can view the discreteness in the template lattice as roughly
equivalent to an ignorance on our part about the location of the
manifold ${\cal S}$ of true gravitational wave signals between the
lattice points.  The maximum distance any correct waveform ${\bf
h}(\bftheta)$ could be away from where we may think it should be (where
our guess is for example obtained by linearly extrapolating from the
nearest points on the lattice) is of order $D_{\rm max}$.  We can
crudely view this ignorance as equivalent to a numerical error $\delta
{\bf h}$ in the templates of magnitude $||\delta {\bf h}||= \sqrt{2
\varepsilon_{\rm grid}} \rho$.  Combining Eqs.~(\ref{Deltadef})
and (\ref{anss}) shows that the loss of information $\delta I_{\rm
source}$ due to the discreteness of the grid should therefore be of
order 
\begin{equation}
\delta I_{\rm source} \sim \rho^2 \left({\rho^2 \over {\cal N}_{\rm
param} + \rho^2}
\right) \, \left( {{\cal N}_{\rm param} \over {\cal N}_{\rm
bins}} \right) \,\, \varepsilon_{\rm grid}.
\label{anss3}
\end{equation}
The grid fineness $\varepsilon_{\rm grid}$ should be chosen to ensure
that $\delta I_{\rm source}$ is small compared to unity, while also
taking into account that the fractional loss in event detection rate
for signal searches due to the coarseness of the grid will be
$\alt 3 \varepsilon_{\rm grid}$; see Sec.~\ref{derivationacc}
above and Refs.~\cite{owen,noteowen}.

\section{CONCLUSIONS}
\label{concl}

Theoretical template waveforms for the merger phase of BBH
coalescences from numerical relativity will be a great aid to the
analysis of detected BBH coalescence events.  A complete
bank of templates could be used to implement a matched filtering
analysis of merger data, which would allow measurements of the
binary's parameters and tests of general relativity in a strong field,
highly dynamic, highly non-spherical regime.  Such matched filtering
may also be possible without a complete bank of templates, if
iterative supercomputer simulations are carried out in tandem with the
data analysis.  A match of the detected waves with
those produced by numerical relativity will be a triumph for the theory of
general relativity and an unambiguous signature of the existence of black
holes.  Qualitative information from representative
supercomputer simulations will also be useful, both as an input to
algorithms for extracting the merger waveform's shape from the noisy
interferometer data stream, and as an aid to interpreting the
observed waveforms and making deductions about the waves' source.

We have derived, using several rather different conceptual starting
points, accuracy requirements that numerical templates must satisfy in
order for them to be useful as data analysis tools.  We first
considered matched filtering signal searches using templates; here the
loss in event rate due to template inaccuracies is simply related to
the degradation in SNR, and leads to a criterion on template accuracy.
Approximately the same criterion is obtained when one demands that the
systematic errors in parameter extraction be small compared to the
detector-noise induced statistical errors.  Finally, we   
quantified the information that is encoded in the merger waveforms using
the mathematical framework of information theory, and deduced how much
of the information is lost due to template inaccuracies or to having
insufficiently many templates.  We deduced approximate requirements that
templates must satisfy (in terms of both accuracy of individual
templates and of the spacing between templates) in order that all of
the waveforms information can be extracted.

The theory of maximum likelihood estimation is a useful starting point
for deriving algorithms for reconstructing the gravitational waveforms
from the noisy interferometer output.  In this paper we have discussed
and derived such algorithms in the contexts of both a single detector
and a network of several detectors;
these algorithms can be tailored to build-in many different kinds of
prior information about the waveforms.

\acknowledgments

We thank Kip Thorne for suggesting this project to us, and for his
invaluable encouragement and detailed comments on the paper.  We thank
David Chernoff for some helpful conversations.  This
research was supported in part by NSF grants PHY--9424337,
PHY--9220644, PHY--9514726, and NASA grant NAGW--2897.  S.\ A.\ H.\
gratefully acknowledges the support of a National Science Foundation
Graduate Fellowship, and \'E.\ F.\ likewise acknowledges the support
of Enrico Fermi and Sloan Foundation fellowships.

\appendix

\section{WAVEFORM RECONSTRUCTION USING A NETWORK OF DETECTORS}
\label{network}

In this appendix we describe how to extend the filtering methods
discussed in Sec.~\ref{bestguess} above from a single detector to a
network of an arbitrary number of detectors.  The underlying principle
is again simply to use the maximum likelihood estimator of the
waveform shape.  We also explain the relationship between our waveform
reconstruction method and the method of G\"ursel and Tinto
\cite{tinto}.  Secs.~\ref{deriveposterior} and \ref{direction} below
overlap somewhat with unpublished analyses by Sam Finn
\cite{FinnRecent}.

We start by establishing some notations for a network of detectors;
these notations and conventions follow those of Appendix A of
Ref.~\cite{cutlerflan94}.  The output of such a network can be
represented as a vector ${\mvec s}(t) =[s_1(t), \ldots , s_{n_d}(t)]$,
where $n_d$ is the number of detectors, and $s_a(t)$ is the strain
amplitude read out from the $a$th detector \cite{notationnote}.  There
will be two contributions to the detector output ${\mvec s}(t)$---the
intrinsic detector network noise ${\mvec n}(t)$ (a vector random
process), and the true gravitational wave signal ${\mvec h}(t)$ :
\begin{equation}
{\mvec s}(t) = {\mvec h}(t) + {\mvec n}(t).
\label{basic0}
\end{equation}
We will assume that the detector network noise is stationary and Gaussian.
In reality the noise will be non-stationary and non-Gaussian, but
understanding the optimal method of waveform reconstruction under our
idealized assumptions is an important first step towards more
sophisticated waveform reconstruction algorithms that incorporate more
information about the nature of the noise.
With this assumption, the statistical properties of the detector
network noise can be described by the auto-correlation matrix
\FL
\begin{equation}
\label{c_n_def1}
C_n(\tau)_{ab} =  \langle n_a(t + \tau) n_b(t) \rangle -
\langle n_a(t + \tau) \rangle \, \langle n_b(t) \rangle, 
\end{equation}
where the angular brackets mean an ensemble average or a time average.
The Fourier transform of the correlation matrix, multiplied by two, is
the power spectral density matrix:
\begin{equation}
\label{s_h_def}
S_h(f)_{ab} = 2 \int_{-\infty}^{\infty} d \tau \, e^{2 \pi i f \tau}
C_n(\tau)_{ab}.
\end{equation}
The off-diagonal elements of this matrix describe the effects of
correlations between the noise sources in the various detectors, while
each diagonal element $S_h(f)_{aa}$ is just the usual power spectral
density of the noise in the $a$th detector.
We assume that the functions $S_h(f)_{ab}$ for $a \ne b$ have been
measured for each pair of detectors.

The Gaussian random process ${\mvec n}(t)$ determines a natural inner
product $\left( \ldots | \ldots \right)$ on the space of functions
${\mvec h}(t)$, which generalizes the inner product (\ref{innerprod})
discussed in the body of the paper in the context of a single detector.
The inner product is defined so that the probability that the
noise takes a specific value ${\mvec n}_0(t)$ is
\begin{equation}
\label{noise_distribution}
p[{\mvec n} = {\mvec n}_0] \, \propto \, e^{- \left( {\mvec n}_0 |
{\mvec n}_0 \right) / 2},
\end{equation} 
and it is given by 
\FL
\begin{equation}
\label{product_def}
\left( {\mvec g} \, | \, {\mvec h} \right) \equiv 4 \, {\rm Re}
\int_0^\infty df 
\,\, {\tilde g}_a(f)^* \left[ {\bf S}_h(f)^{-1} \right]^{ab}  
{\tilde h}_b(f).
\end{equation}
See, {\it e.g.}, Appendix A of Ref.~\cite{cutlerflan94} for more details.  

Turn, now, to the relation between the gravitational wave signal
$h_a(t)$ seen in the $a$th detector, and the two independent
polarization components $h_+(t)$ and $h_\times(t)$ of the
gravitational waves.  Let ${\bf x}_a$ be the position and ${\bf d}_a$
the polarization tensor of the $a$th detector in the detector
network.  By polarization tensor we mean that tensor ${\bf d}_a$ for
which the detector's output $h_a(t)$ is given in terms of the waves'
transverse traceless strain tensor ${\bf h}({\bf x},t)$ by
\begin{equation}
\label{polardef}
h_a(t) = {\bf d}_a : {\bf h}({\bf x}_a,t),
\label{sbd1}
\end{equation}
where the colon denotes a double contraction.  A gravitational wave
burst coming from the direction of a unit vector ${\bf m}$ will have
the form
\begin{equation}
{\bf h}({\bf x},t) = \sum_{A = +,\times} \, h_A(t + {\bf m} \cdot {\bf
x})\,  {\bf e}^A_{\bf m},
\label{sbd2}
\end{equation}
where ${\bf e}^+_{\bf m}$ and ${\bf e}^\times_{\bf m}$ are a basis for
the transverse traceless tensors perpendicular to ${\bf m}$,
normalized according to ${\bf e}^A_{\bf m} : {\bf e}^B_{\bf m} = 2
\delta^{AB}$.  (Note that the notation ${\bf n}$ is typically used
to denote direction to the source; we use instead ${\bf m}$ because we
have denoted by ${\bf n}$ the detector noise.)
Combining Eqs.~(\ref{sbd1}) and
(\ref{sbd2}) and switching from the time domain to the frequency
domain using the convention (\ref{fourierdef}) yields
\begin{equation}
{\tilde h}_a(f) = F_a^A({\bf m}) \, {\tilde h}_A(f) \, e^{- 2 \pi i f
\tau_a({\bf m})},
\label{dependence}
\end{equation}
where the quantities
\begin{equation}
\label{formfactor}
F^A_a({\bf m})  \equiv {\bf e}^A_{\bf m} : {\bf d}_a,
\end{equation}
for $A = +,\times$, are detector beam-pattern functions
for the $a$th detector \cite{300yrs} and $\tau_a({\bf m}) \equiv {\bf m} \cdot
{\bf x}_a$ is the time delay at the $a$th detector relative to the
origin of coordinates.

\subsection{Derivation of posterior probability distribution}
\label{deriveposterior}

We now construct the probability distribution ${\cal P}[{\bf m}, h_+(t),
h_\times(t) | {\mvec s}(t) ]$ for the gravitational waves to be coming
from direction ${\bf m}$ with waveforms $h_+(t)$ and $h_\times(t)$,
given that the output of the detector network is ${\mvec s}(t)$.
Let $p^{(0)}({\bf m})$ and $p^{(0)}[h_A(t)]$ be the prior probability
distributions for the sky position ${\bf m}$ (presumably a uniform
distribution on the unit sphere) and waveform shapes
$h_A(t)$, respectively.  A standard Bayesian analysis along the lines
of that given in Ref.~\cite{finnmeasure} and using
Eq.~(\ref{noise_distribution}) gives 
\FL
\begin{eqnarray}
{\cal P}[{\bf m}, h_A(t) | {\mvec s}(t)] &=& {\cal K} \,
p^{(0)}({\bf m}) \, p^{(0)}[h_A(t)] \nonumber \\
\mbox{} &&\times \, \exp \left[ - \left( {\mvec s} -
{\mvec h} \,|\, {\mvec s} - {\mvec h} \right) /2 \right],
\label{pfundamental}
\end{eqnarray}
where ${\cal K}$ is a normalization constant and ${\mvec h} = (h_1,
\ldots, h_{n_d})$ is understood to be the function of ${\bf m}$ and
$h_A(t)$ given by (the Fourier transform of) Eq.~(\ref{dependence}).

We simplify the expression (\ref{pfundamental}) in two stages.  
First, we reduce the argument of the exponential from a double sum
over detectors to a single sum over detector sites.  In the next few
paragraphs we carry out this reduction, leading to
Eqs.~(\ref{pfundamental3}) and (\ref{pfundamental4}) below.

We assume that each pair of detectors in the detector network comes in
one of two categories: (i) pairs of detectors at the same detector
site, which are oriented the same way, and thus share common detector
beam pattern functions $F_a^A({\bf m})$ (for example the 2 km and 4 km
interferometers at the LIGO Hanford site); or (ii) pairs of detectors
at widely separated sites, for which the detector noise is effectively
uncorrelated.  Under this assumption we can arrange for the matrix
${\bf S}_h(f)$ to have a block diagonal form, with each block
corresponding to a detector site, by choosing a suitable ordering of
detectors in the list $(1, \ldots, n_d)$.  Let us denote the detector
sites by Greek indices $\alpha, \beta, \gamma \ldots$, so that
$\alpha$ runs from $1$ to $n_s$, where $n_s$ is the number of sites.
Let ${\cal D}_\alpha$ be the subset of the list of detectors $(1,
\ldots, n_d)$ containing the detectors at the $\alpha$th site, so that
any sum over detectors can be rewritten
\begin{equation}
\sum_{a=1}^{n_d} = \, \sum_{\alpha=1}^{n_s} \, \sum_{a \, \in {\cal
D}_\alpha}.
\end{equation}
Thus, for example, for a 3 detector network with 2 detectors at the first
site and one at the second, ${\cal D}_1 = \{1,2\}$ and ${\cal D}_2 = \{3\}$.
Let $F_\alpha^A({\bf m})$ denote the common value of the beam pattern
functions (\ref{formfactor}) for all the detectors at site $\alpha$.
Let ${\bf S}_\alpha(f)$ denote the $\alpha$th diagonal sub-block of the matrix
${\bf S}_h(f)$.  Then if we define
\begin{equation}
\Lambda = \left( {\mvec s} -{\mvec h} \,|\, {\mvec s} - {\mvec h} \right),
\label{Lambdadef}
\end{equation}
[the quantity which appears in the exponential in Eq.\
(\ref{pfundamental})], we obtain from Eq.~(\ref{product_def})
\begin{eqnarray}
\Lambda &=& \sum_{\alpha=1}^{n_s} \, 4 \, {\rm Re} \, \int_0^\infty \, df
\, \sum_{a,b \, \in {\cal D}_\alpha} 
\,\, \left[{\tilde s}_a(f)^* - {\tilde h}_a(f)^* \right] \, \nonumber
\\
\mbox{} && \times \left[ {\bf
S}_\alpha(f)^{-1} \right]^{ab} \, \left[{\tilde s}_b(f) - {\tilde
h}_b(f) \right]. 
\label{pfundamental1}
\end{eqnarray}

Next, we note from Eq.~(\ref{dependence}) that the value of ${\tilde
h}_a$ will be the same for all detectors at a given site $\alpha$.  If we 
denote this common value by ${\tilde h}_\alpha$, then we obtain
after some manipulation of Eq.~(\ref{pfundamental1})
\FL
\begin{equation}
\Lambda = \sum_{\alpha=1}^{n_s} \, 4 \, {\rm Re} \, \int_0^\infty \, df
\,\,\left\{ { | {\tilde s}_\alpha(f) - {\tilde h}_\alpha(f) |^2 
\over S_\alpha^{({\rm eff})}(f) } + \Delta_\alpha(f) \right\}.
\label{pfundamental2}
\end{equation}
The meanings of the various symbols in Eq.~(\ref{pfundamental2}) are
as follows.  The quantity $S_\alpha^{({\rm eff})}(f)$ is
defined by
\begin{equation}
{1 \over S_\alpha^{({\rm eff})}(f) } \equiv \sum_{a,b \, \in {\cal D}_\alpha}
\left[ {\bf S}_\alpha(f)^{-1} \right]^{ab},
\label{effnoise}
\end{equation}
and can be interpreted as the effective overall noise spectrum for
site $\alpha$ \cite{noteeff}.  The quantity $s_\alpha$ is given by
\begin{equation}
{\tilde s}_\alpha(f) \equiv S_\alpha^{({\rm eff})}(f) \, \sum_{a,b \,
\in {\cal D}_\alpha} \left[ {\bf S}_\alpha(f)^{-1} \right]^{ab}
{\tilde s}_b(f),
\label{siteoutputdef}
\end{equation}
and is, roughly speaking, the mean output strain amplitude of site
$\alpha$.  Finally,
\begin{eqnarray}
\Delta_\alpha(f) &\equiv& \sum_{a,b \, \in {\cal D}_\alpha} {\tilde
s}_a(f)^* {\tilde s}_b(f) \bigg\{ \left[{\bf
S}_\alpha(f)^{-1}\right]^{ab} \nonumber \\
\mbox{} &&-  S_\alpha^{({\rm eff})}(f) \sum_{c,d \, \in {\cal D}_\alpha}
\left[ {\bf S}_{\alpha}(f)^{-1} \right]^{ac} \,
\left[ {\bf S}_{\alpha}(f)^{-1} \right]^{db}
\bigg\}.
\end{eqnarray}
The quantity $\Delta_\alpha$ is independent of ${\bf m}$ and $h_A(t)$,
and is therefore irrelevant for our purposes; it can be absorbed into
the normalization constant ${\cal K}$ in Eq.~(\ref{pfundamental}).
This unimportance of $\Delta_\alpha$ occurs because we are assuming
that there is some signal present.  However, in situations where one
is trying to assess the probability that some signal (and not just
noise) is present in the outputs of the detector network, the term
$\Delta_\alpha$ is very important.  In effect, it encodes the
discriminating power against noise bursts which is due to the presence
of detectors with different noise spectra at one site ({\it e.g.}, the
2km and 4km interferometers at the LIGO Hanford site).  We drop the
term $\Delta_\alpha$ from now on.

The probability distribution for the waveform shapes and sky direction is
now given by, from Eqs.~(\ref{pfundamental}), (\ref{Lambdadef}) and
(\ref{pfundamental2}), 
\FL
\begin{equation}
{\cal P}[{\bf m}, h_A(t) | {\mvec s}(t)] = {\cal K} \,
p^{(0)}({\bf m}) \, p^{(0)}[h_A(t)] \, e^{- \Lambda^\prime/2},
\label{pfundamental3}
\end{equation}
where
\begin{equation}
\Lambda^\prime = \sum_{\alpha=1}^{n_s} \, 4 \, {\rm Re} \, \int_0^\infty \, df
\,\, { | {\tilde s}_\alpha(f) - {\tilde h}_\alpha(f) |^2 
\over S_\alpha^{({\rm eff})}(f) }.
\label{pfundamental4}
\end{equation}
Finally, we express this probability distribution directly in terms of the
waveforms $h_+(t)$ and $h_\times(t)$ by substituting
Eq.~(\ref{dependence}) into Eq.~(\ref{pfundamental4}), which gives
\begin{eqnarray}
\Lambda^\prime &=& 4 {\rm Re} \int_0^\infty df \bigg\{ \sum_{A,B =
+,\times} \Theta^{AB}(f,{\bf m}) \, \left[ {\tilde h}_A(f)^* - {\tilde
{\hat h}}_A(f)^* \right] \nonumber \\
\mbox{} && \times \, \left[ {\tilde h}_B(f) - {\tilde {\hat h}}_B(f)
\right] + {\cal S}(f,{\bf m}) \bigg\}.
\label{pfundamental5}
\end{eqnarray}
Here
\begin{equation}
\Theta^{AB}(f,{\bf m}) \equiv \sum_{\alpha=1}^{n_s} \, {
F_\alpha^A({\bf m}) F_\alpha^B({\bf m}) \over S_\alpha^{({\rm
eff})}(f) },
\label{Thetadef}
\end{equation}
\begin{equation}
{\tilde {\hat h}}_A(f) \equiv \Theta_{AB}(f,{\bf m}) \,
\sum_{\alpha=1}^{n_s} \,  F_\alpha^B({\bf m}) {\tilde s}_\alpha(f) 
\, e^{2 \pi i f \tau_\alpha({\bf m})},
\label{hathdef}
\end{equation}
where $\Theta_{AB}$ is the inverse matrix to $\Theta^{AB}$, and 
\begin{equation}
{\cal S}(f,{\bf m}) = \sum_\alpha |{\tilde s}_\alpha(f) |^2 -
\Theta^{AB} {\tilde {\hat h}}_A(f)^* {\tilde {\hat h}}_B(f).
\end{equation}

\subsection{Estimating the waveform shapes and the direction to the source}
\label{direction}

Equations (\ref{pfundamental3}) and (\ref{pfundamental5}) constitute
one of the main results of this Appendix, and give the final and
general probability 
distribution for ${\bf m}$ and $h_A(t)$.  In the next few paragraphs
we discuss its implications.  As mentioned at the start of the
appendix, we are primarily interested in situations where the direction ${\bf
m}$ to the source is already known.  However, as an aside, we now
briefly consider the more general context where the direction to the
source as well as the waveform shapes are unknown.

Starting from Eq.~(\ref{pfundamental3}), one could use either maximum
likelihood estimators or so-called Bayes estimators 
\cite{cutlerflan94,schutz1,Davis,cardiff} to determine
``best-guess'' values of ${\bf m}$ and $h_A(t)$.  
Bayes estimators have significant advantages over maximum likelihood
estimators but are typically much more difficult to compute, as
explained in, for example, Appendix A of Ref.~\cite{cutlerflan94}.
The Bayes estimator for the
direction to the source will be given by first integrating
Eq.~(\ref{pfundamental3}) over all waveform shapes, which yields
\begin{equation}
{\cal P}[{\bf m} | {\mvec s}(t)] = {\cal K} p^{(0)}({\bf m}) \, {\cal
D}({\bf m}) \,\,\exp \left[ - 2 \int_0^\infty df \, {\cal S}(f,{\bf m})
\right],
\label{preducedn}
\end{equation}
where ${\cal D}({\bf m})$ is a determinant-type factor that is
produced by integrating over the waveforms $h_A(t)$.  This factor
encodes the information that the detector network has greater
sensitivity in some directions than in others, and that other things
being equal, a signal is more likely to have come from a direction in
which the network is more sensitive.  The Bayes estimator of ${\bf m}$
is now obtained simply by calculating the expected value of ${\bf m}$
with respect to the probability distribution (\ref{preducedn}).  The
simpler, maximum likelihood estimator of ${\bf m}$ is given by
choosing the values of ${\bf m}$ [and of $h_A(t)$] which maximize the
probability distribution (\ref{pfundamental3}), or equivalently by
minimizing the quantity
\begin{equation}
\int_0^\infty df \, {\cal S}(f,{\bf m}).
\label{timedelayinfo}
\end{equation}
Let us denote this value of ${\bf m}$ by ${\bf m}_{\rm ML}({\vec s})$.
Note that the quantity (\ref{timedelayinfo}) encodes all the
information about time delays between the signals detected at the
various detector sites; as is well known, directional information is
obtained primarily through time delay information \cite{schutz1}.

In Ref.~\cite{tinto}, G\"ursel and Tinto suggest a method of
estimating ${\bf m}$ from ${\mvec s}(t)$ for a network of three
detectors.  For white noise and for the special case of one detector
per detector site, the G\"ursel-Tinto estimator is the same as the
maximum likelihood estimator ${\bf m}_{\rm ML}({\vec s})$ just
discussed, with one major modification: in Sec.\ V of Ref.\
\cite{tinto}, G\"ursel and Tinto prescribe discarding those Fourier
components of the data whose SNR is
below a certain threshold as the first stage of calculating their estimator.

Turn, now, to the issue of estimating the waveform shapes $h_+(t)$ and
$h_\times(t)$.  In general situations where both ${\bf m}$ and
$h_A(t)$ are unknown, the best way to proceed in principle would be to
integrate the probability distribution (\ref{pfundamental3}) over all
solid angles ${\bf m}$ to obtain a reduced probability distribution
${\cal P}[h_A(t) | {\mvec s}(t)]$ for the waveform shapes, and to use
this reduced probability distribution to make estimators of $h_A(t)$.
However, such an integration cannot be performed analytically and
would not be easy numerically; in practice simpler estimators will
likely be used.  One such simpler estimator is the maximum likelihood
estimator of $h_A(t)$ obtained from Eq.~(\ref{pfundamental3}).  In the
case of no prior information about the waveform shape when the prior
distribution $p^{(0)}[h_A(t)]$ is very broad, this maximum likelihood
estimator is simply ${\hat h}_A(t)$ evaluated at the value ${\bf
m}_{\rm ML}({\vec s})$ of ${\bf m}$ discussed above.

For BBH mergers, in many cases the direction ${\bf m}$ to the source
will have been measured from the inspiral portion of the waveform, and
thus for the purposes of estimating the merger waveform's shape, ${\bf
m}$ can be regarded as known.  The probability distribution for
$h_A(t)$ given ${\bf m}$ and ${\mvec s}(t)$ is, from
Eq.~(\ref{pfundamental3}),
\begin{equation}
{\cal P}[h_A(t) \,| \, {\bf m}, {\mvec s}(t)] = {\cal K}^\prime \,
p^{(0)}[h_A(t)] \, e^{- \Lambda^{\prime\prime}/2}.
\label{pfundamental6}
\end{equation}
Here ${\cal K}^\prime$ is a normalization constant, and
$\Lambda^{\prime\prime}$ is given by Eq.~(\ref{pfundamental5}) with
the term ${\cal S}(f,{\bf m})$ omitted.  The maximum likelihood
estimator of $h_A(t)$ obtained from this probability distribution in
the limit of no prior information is again just ${\hat h}_A(t)$.  The
formula for the estimator ${\hat h}_A(t)$ given by
Eqs.~(\ref{effnoise}), (\ref{siteoutputdef}), (\ref{Thetadef}) and
(\ref{hathdef}) is one of the key results of this appendix.  It
specifies the best-fit waveform shape as a unique function of the
detector outputs $s_a(t)$ for any network of detectors.

\subsection{Incorporating prior information}
\label{priortoo}

In Sec.\ \ref{bestguess}, we suggested a method of reconstruction of
the merger waveform shape, for a single detector, which incorporated
assumed prior information as to the waveform's properties.  In this
appendix, our discussion so far has neglected all prior information
about the shape of the waveforms $h_+(t)$ and $h_\times(t)$.  We now
discuss waveform estimation for a network of detectors, incorporating
prior information, for fixed sky direction ${\bf m}$.

With a few minor modifications, the entire discussion of
Sec.~\ref{bestguess} can be applied to a network of detectors.  The
required modifications are as follows.  First, the linear space $V$
should be taken to be the space of pairs of waveforms $\{h_+(t),
\,h_\times(t) \}$, suitably discretized, so that the dimension of $V$ is 
$2 T^\prime / \Delta t$.  Second, the inner product
(\ref{discreteinnerprod}) must be replaced by a discrete version of
the inner product 
\FL
\begin{eqnarray}
\left( \{h_+, h_\times \} | \{k_+,k_\times\} \right)
&\equiv& 4 {\rm Re} \, \int_0^\infty df \, \Theta^{AB}(f,{\bf m})
\nonumber \\ \mbox{} && 
\times 
\, {\tilde h}_A(f)^* \, {\tilde k}_B(f),
\label{innerprodnet}
\end{eqnarray}
since the inner product (\ref{innerprodnet}) plays the same role in
the probability distribution (\ref{pfundamental6}) as the inner
product (\ref{discreteinnerprod}) plays in the distribution
(\ref{pdf0}).  Third, the estimated waveforms $\{{\hat h}_+(t), {\hat
h}_\times(t)\}$ 
given by Eq.~(\ref{hathdef}) take the place of the measured waveform
${\bf s}$ in Sec.~\ref{bestguess}, for the same reason.  Fourth, 
the wavelet basis used to specify the prior information must be
replaced by a basis of the form $\{w^+_{ij}(t), 
w^\times_{kl}(t)\}$, where $w^+_{ij}(t)$ is a wavelet basis of the
type discussed in Sec.~\ref{bestguess} for the space of waveforms
$h_+(t)$, and $w^\times_{kl}(t)$ is a similar wavelet basis for the
space of waveforms $h_\times(t)$.  The prior information about, for
example, the assumed duration and bandwidths of the waveforms $h_+(t)$
and $h_\times(t)$ can then be represented exactly as in
Sec.~\ref{bestguess}.  With these modifications, the remainder of the
analyses of Sec.~\ref{bestguess} apply directly to a network of
detectors.  Thus the ``perpendicular projection'' estimator (\ref{bestfit})
and the more general estimator (\ref{bestfitsupervalue})
(corresponding to the more general algorithm described in
Sec.~\ref{extendedmethod}) can both be applied to a network of
detectors.

\subsection{The G\"ursel-Tinto waveform estimator}
\label{GTest}

As mentioned in Sec.~\ref{bestguess} above, G\"ursel and Tinto
have suggested an estimator of the waveforms $h_+(t)$ and
$h_\times(t)$ for networks of three detector sites with one detector
at each site \cite{tinto1}, in the case when the direction ${\bf m}$
to the source is known.  In our notation, the construction of that
estimator can be summarized as follows.  First, assume that the
estimator is some linear combination of the outputs of the independent
detectors corrected for time delays: 
\begin{equation}
{\tilde {\hat h}}^{(GT)}_A(f) = \sum_{\alpha=1}^3 w_A^\alpha({\bf m})
\, \, e^{2 \pi i f \tau_\alpha({\bf m})} \, {\tilde s}_\alpha(f).
\label{GTansatz}
\end{equation}
Here ${\tilde {\hat h}}_A^{(GT)}$ is the G\"ursel-Tinto ansatz for the
estimator, and $w_A^\alpha$ are some arbitrary constants that depend
on ${\bf m}$.  [Since
there is only one detector per site we can neglect the distinction
between the output ${\tilde s}_a(f)$ of an individual detector and the
output ${\tilde s}_\alpha(f)$ of a detector site.]  Next, demand that
for a noise-free signal, the estimator reduces to the true waveforms
$h_A(t)$.  From Eqs.~(\ref{basic0}) and (\ref{dependence}) above, this
requirement is equivalent to the equation 
\begin{equation}
\sum_{\alpha=1}^3 \, w_A^\alpha({\bf m}) \, F_\alpha^B({\bf m}) =
\delta^B_A.
\label{constrainw}
\end{equation}
There is a two dimensional linear space of tensors $w_A^\alpha$ which
satisfy Eq.~(\ref{constrainw}).  Finally, choose $w_A^\alpha$ subject
to Eq.~(\ref{constrainw}) to minimize the expected value with respect
to the noise of the quantity
\begin{equation}
\sum_{A = +,\times} \, \int dt \,\,  | {\hat h}_A^{(GT)}(t) - h_A(t) |^2,
\label{GTinnerprod}
\end{equation}
where ${\hat h}_A^{(GT)}(t)$ is given as a functional of $h_A(t)$ and
the detector noise $n_\alpha(t)$ by Eqs.~(\ref{basic0}),
(\ref{dependence}) and (\ref{GTansatz}).

It is straightforward to show by a calculation using Lagrange
multipliers that the resulting estimator is given by \cite{noteI}
\begin{equation}
{\hat h}_A^{(GT)}(t) = {\hat h}_A(t).
\end{equation}
In other words, the G\"ursel-Tinto estimator coincides with the
maximum likelihood estimators of $h_+(t)$ and $h_\times(t)$ discussed
in this appendix in the case of little prior information.  However,
the estimators discussed here generalize 
the G\"ursel-Tinto estimator by allowing an arbitrary number of
detectors per site [with the effective output and effective noise
spectrum of a site being given by Eqs.~(\ref{siteoutputdef}) and
(\ref{effnoise}) above], by allowing an arbitrary number of sites, and
by allowing one to incorporate prior information about the waveform
shapes.

\section{MEASURES OF INFORMATION}
\label{app_info}

In this appendix we substantiate the claims concerning information
theory made in Sec.~\ref{infosec} of the body of the paper.  First, we
argue that the concept of the ``relative information'' of two PDFs
introduced in Eq.~(\ref{infodef}) does have the interpretation we ascribed
to it: it is the base 2 logarithm of the number of distinguishable
measurement outcomes.  Second, we derive the approximate equations
(\ref{infoapprox}) and (\ref{infoapprox1}).

Consider first the issue of ascribing to any measurement process a
``number of bits of information gained'' from that process, which
corresponds to the base 2 logarithm of the number of distinguishable
possible outcomes of the measurement.  If $p^{(0)}({\bf x})$ is the
PDF for the measured quantities ${\bf x} = (x^1, \ldots, x^n)$ before
the measurement, and $p({\bf x})$ is the corresponding PDF after the
measurement, then the relative information of these two PDFs is
defined to be
\begin{equation}
I = \int d^n {\bf x} \, p({\bf x}) \, \log_2 \left[ {p({\bf x}) \over
p^{(0)}({\bf x}) }\right].
\label{rel_info}
\end{equation}
In simple examples, it is easy to see that the quantity
(\ref{rel_info}) reduces to the number of bits of information gained
in the measurement.  For instance, if ${\bf x} = (x^1)$ and the prior
PDF $p^{(0)}$ constrains $x^1$ to lie in some range of size $X$, and
if after the measurement $x^1$ is constrained to lie in a small
interval of size $\Delta x$, then $I \approx \log_2(X / \Delta x)$, as one
would expect.  In addition, the quantity (\ref{rel_info}) has the
desirable feature that it is coordinate independent, {\it i.e.}, that
the same answer is obtained when one makes a nonlinear coordinate
transformation on the manifold parameterized by $(x^1, \ldots, x^n)$
before evaluating the quantity (\ref{rel_info}).  For these reasons,
in any measurement process, the quantity (\ref{rel_info}) can be
interpreted as the number of bits of information gained
\cite{coverthomas}.

\subsection{Explicit formula for the total information}

As a foundation for deriving the approximate formula
(\ref{infoapprox}), we derive in this subsection an explicit
formula [Eq.~(\ref{pfinalans})] for the total information gain
(\ref{infodef}) in a gravitational wave measurement. 
We shall use a basis of $V$ where the matrix (\ref{Gammadef}) is 
unity, and for ease of notation we shall denote by ${\cal N}$ the
quantity which was denoted by ${\cal N}_{\rm bins}$ in the body of the
paper. 

First, we assume that the prior PDF $p^{(0)}({\bf h})$ appearing in
Eq.~(\ref{standardb}) is a function only of $h = ||{\bf h}||$. 
In other words, all directions in the vector space $V$ are taken to
be, {\it a priori}, equally likely, when one measures distances and
angles with the inner product (\ref{discreteinnerprod}).  It would be
more realistic to make such an assumption with respect to a
noise-independent inner product like $(h_1\, | \, h_2) \equiv \int
dt h_1(t) h_2(t)$, but if the noise spectrum $S_h(f)$ does not vary
too rapidly within the bandwidth of interest, the distinction is
not too important and our assumption will be fairly realistic.
We write the prior PDF as \cite{caveat1}
\begin{eqnarray}
p^{(0)}({\bf h}) \,  d^{\cal N} h &=& {2 \pi^{{\cal N}/2} \over \Gamma({\cal
N}/2)} \,\, h^{{\cal N}-1} \,\, p^{(0)}(h) dh \nonumber \\
\mbox{} &\equiv& {\bar p}^{(0)}(h) \, dh.
\label{prior00}
\end{eqnarray}
The quantity ${\bar p}^{(0)}(h) \, dh$ is the prior probability that the signal
${\bf h}$ will have an SNR $||{\bf h}||$ between $h$ and $h + dh$.
The exact form of the PDF ${\bar p}^{(0)}(h)$ will not be too important
for our calculations below.  A moderately realistic choice is ${\bar
p}^{(0)}(h) \propto 1/h^3$ with a cutoff at some $h_1 \ll 1$.  Note
however that the choice $p^{(0)}({\bf h}) = 1$ corresponding to ${\bar
p}^{(0)}(h) \propto h^{{\cal N}-1}$ is very unrealistic.  Below we
shall assume that ${\bar p}^{(0)}(h)$ is independent of ${\cal N}$.

We next write Eq.~(\ref{standardb}) in a more explicit form. 
Without loss of generality we can take
\begin{equation}
{\bf s} = (s^1, \ldots, s^{\cal N}) = (s, 0, \ldots, 0),
\end{equation}
where $s = \rho({\bf s}) = || {\bf s} ||$.
Then, writing $( {\bf s} | {\bf h}) = s h \cos \theta$ and
using the useful identity
\begin{equation}
d^{\cal N} h  = {2 \pi^{({\cal N}-1)/2} \over \Gamma\left[ ({\cal
N}-1)/2 \right]} \, \sin(\theta)^{{\cal N}-2} \, h^{{\cal N}-1} \, d \theta
\, d h,
\label{identity}
\end{equation}
we can write
\begin{eqnarray}
p({\bf h}\,|\,{\bf s}) \, d^{\cal N}h &=&  {\cal K}_1 \, {\bar p}^{(0)}(h) \,
\sin(\theta)^{{\cal N}-2} \nonumber \\
\mbox{} && \times \, \exp \left[ - {1 \over 2} \left(s^2 + h^2 - 2
s h \cos  \theta \right) \right] \,\, dh \, d\theta,
\label{pp2}
\end{eqnarray}
where ${\cal K}_1$ is a constant.  If we define the function $F_{\cal
N}(x)$ by 
\begin{equation}
F_{\cal N}(x) \equiv {1 \over 2} \int_0^\pi d\theta \,
\sin(\theta)^{{\cal N}-2} \,\, e^{x \cos \theta},
\label{Fndef}
\end{equation}
then the constant ${\cal K}_1$ is determined by the normalization condition 
\begin{equation}
1 = 2 {\cal K}_1 \int_0^\infty dh \, 
e^{- (s^2 + h^2)/2} \, F_{\cal N}(s h) \, {\bar p}^{(0)}(h).
\label{K1def}
\end{equation}

We next calculate the PDF $p[{\bf h} \,|\, \rho({\bf s})]$ appearing
in the denominator in Eq.~(\ref{infodef}).  From Bayes's theorem, this
PDF is  given by
\begin{equation}
p[{\bf h} \,|\, \rho({\bf s})] = {\cal K} \, p^{(0)}({\bf h}) \,
p[\rho({\bf s}) \, | \, {\bf h} ],
\label{pp2a}
\end{equation}
where $p[\rho({\bf s}) \, | \, {\bf h} ]$ is the PDF for $\rho({\bf
s})$ given that the gravitational wave signal is ${\bf h}$, and ${\cal
K}$ is a normalization constant.  Using the fact that $p({\bf s} \,|\,
{\bf h}) \propto \exp \left[ - ({\bf s} - {\bf h})^2 \right]$, we find
using Eq.~(\ref{identity}) that
\begin{eqnarray}
p({\bf s}\,|\,{\bf h}) \, d^{\cal N} s &=&  {2^{1 - {\cal N}/2} \over
\sqrt{\pi}  
\Gamma\left[({\cal N}-1)/2\right]  } \,\, \sin(\theta)^{{\cal N}-2} \,\,
s^{{\cal N}-1} \nonumber \\
\mbox{} && \times \, \exp \left[ - {1 \over 2} \left(s^2 + h^2 - 2 s h \cos
\theta \right) \right] \,\, ds \, d\theta.
\label{pp3}
\end{eqnarray}
Integrating over $\theta$ 
now yields from Eq.~(\ref{Fndef})
\begin{equation}
p[\rho({\bf s})=s \, | \, {\bf h} ] \, ds \, \propto \, s^{{\cal
N}-1} e^{- (s^2 + h^2)/2} \, F_{\cal N}(s h) 
\, ds.
\label{pp4}
\end{equation}
Now combining Eqs.~(\ref{identity}), (\ref{pp2a}), and (\ref{pp4})
yields 
\begin{eqnarray}
p[{\bf h}\,|\,\rho({\bf s})] \, d^{\cal N}h &=&  
{\cal K}_2 \,\, {\bar p}^{(0)}(h) \,\, e^{-(\rho({\bf s})^2 + h^2)/2} \,
F_{\cal N}[\rho({\bf s}) h]
\nonumber \\
\mbox{} && \times \,\, \sin(\theta)^{{\cal N}-2} \,\, dh \, d\theta,
\label{pp5}
\end{eqnarray}
where from Eq.~(\ref{K1def}) the normalization constant is given by
\begin{equation}
{\cal K}_2 = {2 \Gamma({\cal N}/2) \over \sqrt{\pi} \Gamma[({\cal
N}-1)/2] } \, {\cal K}_1.
\label{K2def}
\end{equation}

We can now calculate the information $I_{\rm total}$ by combining
Eqs.~(\ref{infodef}), (\ref{pp2}), (\ref{Fndef}), (\ref{pp5}), and
(\ref{K2def}).  The result is
\begin{eqnarray}
I_{\rm total}[\rho({\bf s}),{\cal N}] &=& - \int_0^\infty dh \,
p^{(1)}(h) \, G_{\cal N}[\rho({\bf s}) h] \nonumber \\
\mbox{} &&- \log_2 \left[ {2
\Gamma({\cal N}/2) \over 
\sqrt{\pi} \Gamma[({\cal N}-1)/2] } \right],
\label{pfinalans}
\end{eqnarray}
where 
\begin{equation}
G_{\cal N}(x) \equiv {x F_{\cal N}^\prime(x) \over \ln 2 \, F_{\cal
N}(x)} - \log_2 F_{\cal N}(x),
\label{GNdef}
\end{equation}
and
\begin{equation}
p^{(1)}(h) \equiv 2 {\cal K}_1 \, {\bar p}^{(0)}(h) \,\,
e^{-(\rho({\bf s})^2 + h^2)/2} \,\,  F_{\cal N}[\rho({\bf s}) h].
\label{p1def}
\end{equation}
Equations (\ref{K1def}), (\ref{Fndef}), and (\ref{pfinalans}) --
(\ref{p1def}) now define explicitly the total information $I_{\rm
total}$ as a function of the parameters $\rho({\bf s})$ and ${\cal N}$
and of the prior PDF ${\bar p}^{(0)}(h)$.

\subsection{Approximate formula for the total information}

We now derive the approximate formula (\ref{infoapprox}) for the total
information.  Let $\rho^2_b = \rho({\bf s})^2 / {\cal N}$; we will
consider the limit of large $\rho({\bf s})$ and ${\cal N}$ but fixed
$\rho_b$.  Our analysis will divide into two cases, depending on whether
$\rho_b >1 $ or $\rho_b \le 1$.  Let us first consider the case
$\rho_b >1$. In the large ${\cal N}$ limit the result for $\rho_b > 1$
will be independent of the prior PDF ${\bar p}^{(0)}(h)$, which we
assume has no dependence on ${\cal N}$.

The first term in Eq.~(\ref{pfinalans}) is the expected value $\langle
G_{\cal N}[\rho({\bf s}) h] \rangle$ of $G_{\cal N}[\rho({\bf s}) h]$
with respect to the PDF (\ref{p1def}).  If we change the variable of
integration in this term from $h$ to $u = h / \sqrt{{\cal N}}$, we
find
\begin{eqnarray}
\langle \, G_{\cal N}[\rho({\bf s}) h] \, \rangle \,\, &\propto & \,\,
\int_0^\infty du \,\, {\bar p}^{(0)}(\sqrt{{\cal N}} u) \,\, e^{-
{\cal N}(\rho_b^2 + u^2)/2} 
\nonumber \\
\mbox{} && \times \,\, F_{\cal N}({\cal N} \rho_b u) \,\,G_{\cal
N}({\cal N} \rho_b u). 
\label{hardterm}
\end{eqnarray}
From Eq.~(\ref{Fndef}) it is straightforward to show that in the limit
of large ${\cal N}$, 
\begin{equation}
F_{\cal N}({\cal N} z) \approx {1 \over 2} e^{{\cal N} q(\theta_c) }
\,\, \sqrt{{ 2 \pi \over {\cal N} |q''(\theta_c)| }},
\label{Fnapprox}
\end{equation}
for fixed $z$.  Here $q(\theta)$ is the function
\begin{equation}
q(\theta) = z \cos \theta + \ln \sin \theta,
\label{qfndef}
\end{equation}
and $\theta_c = \theta_c(z)$ is the value of $\theta$ which maximizes
the function $q(\theta)$, given implicitly by
\begin{equation}
z \sin^2 \theta_c = \cos \theta_c.
\label{thetacdef}
\end{equation}
We similarly find that
\begin{equation}
F_{\cal N}^\prime({\cal N} z) \approx {1 \over 2} e^{{\cal N} q(\theta_c) }
\,\, \sqrt{{ 2 \pi \over {\cal N} |q''(\theta_c)| }} \, \cos \theta_c.
\label{Fnapprox1}
\end{equation}
It is legitimate to use the approximations (\ref{Fnapprox}) and
(\ref{Fnapprox1}) in the integral (\ref{hardterm}) since the value
$u_{\rm max}({\cal N},\rho_b)$ of $u$ at which the PDF $p^{(1)}({\cal
N}\rho_b u)$ is a maximum approaches at large ${\cal N}$ a constant
$u_{\rm max}(\rho_b)$ which is independent of ${\cal N}$, as we show
below.

Inserting the approximation (\ref{Fnapprox}) into Eq.~(\ref{hardterm})
and identifying $z = \rho_b u$, we find that the PDF (\ref{p1def}) is
proportional to
\begin{equation}
\exp\left[ {\cal N} {\cal Q}(u) + O(1) \right],
\label{pdfapp}
\end{equation}
where
\begin{equation}
{\cal Q}(u) = - {1 \over 2}(\rho_b^2 + u^2) + q(\theta_c)
\label{calQ}
\end{equation}
and $\theta_c = \theta_c(z) = \theta_c(\rho_b u)$.  From Eqs.~(\ref{qfndef})
and (\ref{thetacdef}) it can be shown
that the function (\ref{calQ}) has a local maximum at
\begin{equation}
u_{\rm max} = \sqrt{\rho_b^2-1}
\label{umaxdef}
\end{equation}
at which point $\theta_c$ is given by $\sin \theta_c = 1/\rho_b$.  The
form of the PDF (\ref{pdfapp}) now shows that at large ${\cal N}$,
\begin{equation}
\langle G_{\cal N}({\cal N} \rho_b u) \rangle \approx  \,\, 
G_{\cal N}({\cal N} \rho_b u_{\rm max}).
\label{pp9}
\end{equation}
Finally, if we combine Eqs.~(\ref{pfinalans}),
(\ref{Fnapprox})--(\ref{Fnapprox1}), (\ref{umaxdef}) and (\ref{pp9})
and use Stirling's formula to approximate the Gamma functions, we
obtain Eq.~(\ref{infoapprox0}).

Turn, next, to the case $\rho_b < 1$.  In this case the function
${\cal Q}$ does not have a local maximum, and the dominant
contribution to the integral (\ref{hardterm}) at large ${\cal N}$
comes from $h \sim O(1)$ (rather than from $h \sim \sqrt{{\cal N}}$, $u
\sim O(1)$ as was the case above).  From Eq.~(\ref{Fndef}) we obtain
the approximations
\begin{equation}
F_{\cal N}(\sqrt{{\cal N}} w) = \sqrt{ {\pi \over 2 {\cal N}}} \, e^{w^2/2}
\, \left[ 1 + O(1/\sqrt{{\cal N}}) \right]
\label{Fn3}
\end{equation}
and
\begin{equation}
F_{\cal N}^\prime({\sqrt {\cal N}} w) = \sqrt{ {\pi \over 2 }} \, {w \over
{\cal N}} \, e^{w^2/2} \, \left[ 1 + O(1 / \sqrt{{\cal N}})
\right],
\label{Fn4}
\end{equation}
which are valid for fixed $w$ at large ${\cal N}$.  Using
Eqs.~(\ref{Fn3}), (\ref{Fn4}), and (\ref{pfinalans}) -- (\ref{p1def}),
and using Stirling's formula again we find that
\FL
\begin{equation}
I_{\rm total} \approx {1\over2} \rho_b^2 \,\,
{ \int_0^\infty dh \, {\bar p}^{(0)}(h) \, \exp \left[-(1 - \rho_b^2)
h^2/2 \right]
\,h^2 \over \int_0^\infty dh \, {\bar p}^{(0)}(h) \, \exp \left[-(1 -
\rho_b^2) h^2/2 \right]}.
\end{equation}
For simplicity we now take ${\bar p}^{(0)}(h)$ to be a Gaussian
centered at zero with width $h_{\rm prior}^2$; this yields
\begin{equation}
I_{\rm total} \approx {1 \over 2} \left[ { \rho_b^2 h_{\rm prior}^2
\over 1 + (1 - \rho_b^2) h_{\rm prior}^2 } \right].
\label{I4}
\end{equation}

From Eq.~(\ref{magnitude3}), the parameter $\rho_b$ is given by
\begin{equation}
\rho_b^2 = 1 + {\rho^2 \over {\cal N}_{\rm bins}} \pm {1 \over
{\sqrt{\cal N}_{\rm bins}}},
\end{equation}
where the last term denotes the rms magnitude of the statistical
fluctuations.  Since we are assuming that $\rho_b < 1$, it follows
that $\rho_b^2 \approx 1 - 1/
{\sqrt{\cal N}_{\rm bins}}$, and therefore we obtain from
Eq.~(\ref{I4}) that
\begin{equation}
I_{\rm total} \approx {1 \over 2} {\rm min}\left[ h_{\rm prior}^2,
\sqrt{{\cal N}_{\rm bins}} \right].
\end{equation}
Thus, if $h_{\rm prior} \alt 1$, then the total information gain is
$\alt 1$ also.

\subsection{Approximate formula for the source information}

We now turn to a discussion of the approximate formula
(\ref{infoapprox1}) for the information (\ref{infodef1}) obtained
about the source of the gravitational waves.  In general, the measure
of information (\ref{infodef1}) depends in a complex way on the prior
PDF $p^{(0)}({\bf h})$, and on how the waveform ${\bf h}(\bftheta)$
depends on the source parameters $\bftheta$.  We can evaluate the
information $I_{\rm source}$ explicitly in the simple and unrealistic
model where the dependence on the source parameters $\bftheta$ is
linear and where there is little prior information.  In this case the
manifold of possible signals is a linear subspace (with dimension
${\cal N}_{\rm param}$) of the linear space of all possible signals
(which has dimension ${\cal N}$).  The integral (\ref{infodef1}) then
reduces to an integral analogous to (\ref{infodef}), and we obtain the
formula (\ref{infoapprox1}) in the same way as we obtained
Eq.~(\ref{infoapprox}).  The result (\ref{infoapprox1}) is clearly a
very crude approximation, as the true manifold of merger signals is
very curved and nonlinear.  Nevertheless, it seems likely that the
formula (\ref{infoapprox1}) will be valid for some effective number of
parameters ${\cal N}_{\rm param}$ that is not too much different from
the true number of parameters on which the waveform depends.

\section{EXPECTED VALUE OF CORRELATION COEFFICIENT}
\label{correl}

In this appendix we derive the formula (\ref{cexpected}) for the
expected value of the correlation coefficient (\ref{cdef}).  We start
by deriving the following general result.  Let ${\bf n} = (n^1,
\ldots, n^{\cal N})$ be a Gaussian random variable with $\langle {\bf
n} \rangle = 0$ and $\langle n^i \, n^j \rangle = \Sigma^{ij}$.  Let
${\bf h} = (h^1, \ldots , h^{\cal N})$ be a fixed vector, 
and define the random variable ${\cal C}$ by 
\begin{equation}
{\cal C} = { {\bf h} \cdot {\bf \Sigma}^{-1} \cdot( {\bf h} + {\bf n}) \over
\sqrt{{\bf h} \cdot {\bf \Sigma}^{-1} \cdot {\bf h}} \sqrt{
( {\bf h} + {\bf n}) \cdot {\bf \Sigma}^{-1} \cdot ( {\bf h} + {\bf n})}}.
\label{cdefappendix}
\end{equation}
Then, in the regime ${\cal N} \gg 1$
and $\rho \gg 1$, where $\rho^2 \equiv {\bf h} \cdot {\bf \Sigma}^{-1}
\cdot {\bf h}$, we have
\begin{equation}
\left< {\cal C} \right> = {1 \over \sqrt{ 1 + {\cal N} / \rho^2}} \,
\left[ 1 + O\left( {1 \over \rho^2},\, {1 \over {\cal N}} \, \right)
\right]. 
\label{cans}
\end{equation}

Equation (\ref{cexpected}) can be obtained from Eq.~(\ref{cans}) 
as follows.  From Eq.~(\ref{bestfit}), the vector ${\bf h}_{\rm
best-fit}$ can be written as 
\begin{equation}
{\bf h}_{\rm best-fit} = {\bf h} + {\bf n}_\parallel
\end{equation}
where ${\bf n}_\parallel$ denotes the component of the noise ${\bf n}$
in the space $U$.  Thus, the vectors ${\bf h}$ and ${\bf h}_{\rm
best-fit}$ which appear in Eq.~(\ref{cdef}) both lie in the space $U$
of dimension ${\cal N}_{\rm bins}$ (although both nominally lie in the
larger space $V$ of dimension ${\cal N}_{\rm bins}^\prime$).  
Now, identifying ${\cal N}$ and ${\cal N}_{\rm
bins}$, we see that the quantities (\ref{cdef}) and
(\ref{cdefappendix}) coincide, and the result (\ref{cexpected})
follows.

We now now turn to the derivation of Eq.~(\ref{cans}).  First, make a
linear change of variables to make $\Sigma_{ij} = \delta_{ij}$.  (The
results obtained at the end can be generalized to non-unit ${\bf
\Sigma}$ by inspection.)  We want to evaluate
\begin{equation}
\langle {\cal C} \rangle = \int dn^1 \ldots dn^{\cal N} \  p({\bf n})
{\cal C}({\bf n}), 
\end{equation}
where ${\cal C}({\bf n})$ is given by Eq.~(\ref{cdefappendix}).  
The quantity ${\cal C}({\bf n})$ depends on 
${\bf n}$ only through the combinations
\begin{equation}
\alpha \equiv {\bf h} \cdot {\bf n} = \sum_i \, h_i  n^i
\label{alphadef}
\end{equation}
and
\begin{equation}
\beta \equiv {\bf n} \cdot {\bf n} = \sum_{i=1}^{\cal N} \, (n^i)^2.
\label{betadef}
\end{equation}
Hence
\begin{equation}
\langle {\cal C} \rangle = \int d\alpha \int d\beta \,\, p(\alpha,\beta) \,
{\cal C}(\alpha, \beta),
\label{ansl1}
\end{equation}
where
\begin{equation}
{\cal C}(\alpha,\beta) = {\rho^2 + \alpha \over \rho \sqrt{ \rho^2 + 2 \alpha
+ \beta}}.
\label{c1def}
\end{equation}

The probability distribution $p(\alpha,\beta)$ can be approximately
evaluated in the following way.  We have
\begin{equation}
p(\alpha,\beta) = p(\beta | \alpha) \, p(\alpha).
\end{equation}
Here $p(\beta | \alpha)$ is the distribution for $\beta$ given a value
of $\alpha$, and $p(\alpha)$ is from Eq.~(\ref{alphadef}) a
Gaussian with zero mean and variance $\rho^2$:
\begin{equation}
p(\alpha) = {1 \over \sqrt{2 \pi} \rho} \exp \left\{ - \alpha^2 / (2
\rho^2) \right\}. 
\end{equation}
We introduce the notation 
\begin{equation}
\langle \ldots \rangle_\alpha = \int \, \ldots p(\beta | \alpha) d\beta,
\end{equation}
and define $(\Delta \beta)_\alpha^2 = \langle \beta^2 \rangle_\alpha -
\langle \beta \rangle_\alpha^2$.  The distribution $p(\beta | \alpha)$
can be treated as being approximately Gaussian in the regime where
$(\Delta \beta)_\alpha \ll \langle \beta \rangle_\alpha$, which we show below
is the case when $\rho^2 \gg 1$ and ${\cal N} \gg 1$.  Hence we need
only evaluate $\langle \beta \rangle_\alpha$ and $(\Delta \beta)_\alpha$.

Without loss of generality we can write ${\bf h} = (\rho,0,\ldots,0)$, so
that from Eq.~(\ref{alphadef}), $\alpha = \rho n^1$.  Similarly
Eq.~(\ref{betadef}) gives
\begin{equation}
\beta = {\alpha^2 \over  \rho^2} + 
\sum_{i=2}^{{\cal N}} \,  (n^i)^2.
\label{beta2}
\end{equation}
Using the fact that the $n^i$ are independent normally distributed
random variables, it follows from Eq.~(\ref{beta2}) that 
\begin{equation}
\left< \beta \right>_\alpha = {\alpha^2 \over \rho^2} + {\cal N} -1,
\end{equation}
and similarly we find
\begin{equation}
(\Delta \beta)_\alpha^2 = 2 ({\cal N}-1).
\end{equation}
Now the integral (\ref{ansl1}) will be dominated by contributions from the
regime $\alpha \alt ({\rm a\ few}) \times \rho$.  In this regime, we have
\begin{equation}
{(\Delta \beta)_\alpha \over \langle \beta \rangle_\alpha} \sim 
{1 \over \sqrt{{\cal N}}} \ll 1,
\end{equation}
which justifies our treating the PDF $p(\beta | \alpha)$ as Gaussian.

Combining these results we find
\FL
\begin{equation}
p(\alpha,\beta) \approx {1 \over 2 \pi \alpha (\Delta \beta)_\alpha}
\,  \exp \left\{ - {\alpha^2 \over 2 a^2} - { 
\left( \beta - \langle \beta \rangle_\alpha \right)^2 \over 2
(\Delta \beta)_\alpha^2 }\right\}.
\end{equation}
Inserting this distribution into Eq.~(\ref{ansl1}),
using Eq.~(\ref{c1def}) and expanding to second order in $\alpha$ gives
\begin{eqnarray}
\langle {\cal C} \rangle &=& \int d \alpha \, p(\alpha) \,\,
{\cal C}[\alpha,\langle \beta 
\rangle_\alpha] \,\, \left[ 1 + O(1/{\cal N}) \right]
\nonumber \\
& = & \int d\alpha \, p(\alpha) \, {1 + {\alpha / \rho^2} \over
\sqrt{ (1 + \alpha /\rho^2)^2 + ({\cal N}-1) / \rho^2 
}} \nonumber \\
\mbox{} && \ \  \times \left[ 1 + O(1/{\cal N}) \right]
\nonumber \\
& = &{1 \over \sqrt{ 1 + {\cal N} / \rho^2}} \, \left[ 1 +
O\left( {1 \over \rho^2},\, {1 \over {\cal N}} \right) \right],
\label{cans1}
\end{eqnarray}
as required.


\begin{references}
\bibitem{ordercount} By post-$n$-Newtonian order, we mean
	$O[(v/c)^{2n}]$ in secular terms and $O[(v/c)^{5+2n}]$ in
	dissipative, radiation reaction terms ($2n$ orders beyond the
	leading quadrupole radiation reaction terms).  This is the
	standard counting method for post-Newtonian waveforms.
\bibitem{rangenote} When we say that a source is visible to a distance
	$D$, we mean that the rate of detection of events is roughly the
	same as the rate of occurrence of events within a sphere of radius
	$D$.  Some events within this sphere will be missed, and some
	event outside this sphere will be detected, due to beaming and
	orientation effects in the emission of gravitational waves.
	Rare, optimally oriented sources will be visible out to several
	times $D$.
\bibitem{noterate}
	The factor of $\sim 10$ increase in event rate, which is quoted
	in Sec.~I.E.~of Ref.~\cite{paperI}, comes from a
	combination of a factor $\sim 40$ increase in event rate for matched
	filtering [the cube of the ratio (merger SNR)/ max(ringdown
	SNR,inspiral SNR), taken from Figs. 4, 5 and 6 of Ref.~\cite{paperI}],
	combined with the fact that noise-monitoring searches perform a factor
	$\sim 4$ worse in event rate than matched filtering searches (the
	estimate of ${\cal R}$ in Sec. VI.B of Ref.~\cite{paperI}).
	This estimate of ${\cal R}$ assumes that we know roughly the range of
	frequency bandwidths and temporal durations of merger waves from
	representative supercomputer simulations, in that the parameter
	${\cal N}_{\rm bins}$ is assumed to have the value $60$.  If,
	on the other hand, we are very ignorant as to the range of
	frequency bandwidths  
	and durations of the merger waves, a more appropriate,
	conservative value of ${\cal N}_{\rm bins}$ for the search
	process might be $\sim 300$, which would yield instead ${\cal R}
	\sim 10$ and a corresponding overall gain in event rate of
	noise-monitoring searches for merger waves over matched filtering
	searches for inspiral/ringdown waves of $\sim 4$.
\bibitem{twosided} Our convention for the normalization of the
	spectral noise density $S_h(f)$ is such that
	$$
	\langle n(t)^2 \rangle = \int_0^\infty \,S_h(f) \, df.
	$$
	(``One-sided spectral noise density'').  This determines the constant
	$4$ appearing in Eq.~(\ref{innerprod}).
\bibitem{notedetect} The optimal thresholding criterion for detection 
	will presumably involve an analysis of the combined data from all the
	detectors after candidate events have been
	identified by optimal filtering at each detector separately.
\bibitem{insptmplcount} The required number of inspiral templates
	depends very 
	strongly on the assumed minimum mass of the bodies in the binaries
	searched for; it is practically independent of the maximum mass
	\cite{owen}.  A given search will look for all types of binary
	systems (neutron star-neutron star, black hole-neutron star, black
	hole-black hole), and so the minimum mass (and hence the number of
	search templates) is set by the assumed minimum neutron star mass.
	The required number of templates quoted are the results from
	Ref.~\cite{owen} as the minimum mass varies from $0.2 M_\odot$
	to $1 M_\odot$.
\bibitem{note2} The accuracy with which the various
	waveform parameters will be measurable is currently known only to
	within factors of order $\sim 2$.  Statistical errors in
	maximum likelihood estimators have been calculated in the high
	signal-to-noise approximation
	\cite{finn2,cutlerflan94,poissonwill,krolaknew}, and also more
	accurately using Monte-Carlo simulations \cite{montecarlo}.
	However, ultimately in the data analysis one should use
	estimators (for example Bayes estimators) which perform better
	than maximum likelihood estimators
	\cite{Davis,schutz1,cutlerflan94,cardiff}.  
	In addition, all analyses to date have neglected the
	modulation of the waveform due to the spin-induced precession
	of the orbital plane, an effect which will be important for
	BBHs with rapidly spinning constituents \cite{harisspin,kidder}.
	Thus the true
	parameter-measurement accuracies are still somewhat uncertain.
	The mass-measurement accuracies we have quoted were obtained
	as follow:  We used the result from 
	result from Ref.~\cite{poissonwill} (in the high SNR approximation)
	that $\Delta\eta/\eta=1.5$ 
	for two $10 \,M_\odot$ black holes with $S/N=10$, where $\eta
	= \mu/M$, $M$ is the total mass and $\mu$ the reduced mass of
	the system.  This implies that
	$\Delta\mu/\mu = 0.6$.  Scaling to $S/N \sim 100$ gives
	$\Delta\mu/\mu = 0.06$.  Finally, the 1$\sigma$ confidence limits
	for the two masses in the equal mass case are approximately
	$m_0 \le m_1 \le m_0 f_+(\Delta\mu/\mu)$ and
	$m_0 f_-(\Delta\mu/\mu) \le m_2 \le m_0$, where $m_0$ is the true
	mass and
	$$
	f_\pm(x) = (1-x)^{-3/2} \left[1 \pm \sqrt{1-(1-x)^{5/2}}\right];
	$$
	see Ref.~\cite{cutlerflan94}.  The other accuracies quoted
	were taken from Refs.~\cite{cutlerflan94} and \cite{montecarlo}.
\bibitem{ringnote2} The fraction of events with ringdown SNR greater than
	some threshold $\rho_{\rm ring}$ is $\sim \rho_{\rm insp}^3
	/(\alpha \, \rho_{\rm ring})^3$, where $\rho_{\rm insp}$ is
	the detection 
	threshold for inspiral SNRs, and $\alpha$ is the ratio between the
	actual inspiral and ringdown SNRs.  This assumes that all events are
	detected via their inspiral signals and thus have inspiral SNRs larger
	than $\rho_{\rm insp}$.  Equations (5.4) and (5.5) of
	Ref.~\cite{paperI} imply that $\alpha \sim 30$ for initial and
	advanced LIGO interferometers for $M \alt 50 \, M_\odot$.
	Taking $\rho_{\rm ring} = 1$ and 
	using the typically discussed inspiral SNR threshold of
	$\rho_{\rm insp}=6$ \cite{prl}, one obtains that the fraction
	is about 1\%.
\bibitem{noteF} When a complete set of theoretical templates is
	available so that matched filtering can be carried out,
	reconstructing the ``best-fit'' waveform is of course
	trivial: it is just the template which maximizes the signal to
	noise ratio.
\bibitem{note20}
	As we show in Sec.\ \ref{bestguess}, in the lack of any
	information about the merger signal other than its expected
	bandwidth, this band-pass filtered detector output is in fact
	the maximum likelihood estimator of the merger waveform.
\bibitem{eg} For example, one can first truncate the signal in the
	time domain, then perform a finite Fourier transform, and then
	throw away the Fourier coefficients outside the frequency
	range of interest.  Or, one can first band-pass filter the data
	stream, and then truncate in the time domain.
\bibitem{notec} One could define an alternative correlation
	coefficient by not using the noise inner product
	(\ref{discreteinnerprod}) but instead a different inner
	product.  An example of such an alternative correlation
	coefficient is 
	$$
	{\cal C}^\prime = {\int_{-T/2}^{T/2} dt \, h(t) h_{\rm best-fit}(t)
	\over \sqrt{ \int_{-T/2}^{T/2} h(t)^2 \, dt }\
	\sqrt{\int_{-T/2}^{T/2} h_{\rm best-fit}(t)^2 \, dt } }.
	$$
	When aliasing effects can be neglected, it can be shown using the
	method of appendix \ref{correl} that the
	expected value $\langle {\cal C}^\prime \rangle$ of this
	correlation 
	coefficient is approximately given by Eq.~(\ref{cexpected}), but with
	$\rho_{\rm bin}$ now given by
	$$
	\rho_{\rm bin}^2 = {2 \int_{\Delta f} df \, |{\tilde h}(f) |^2 \over T
	\int_{\Delta f} df \, S_h(f) }.
	$$
\bibitem{note3} Although we discuss only Wiener optimal filtering, there
	are more sophisticated nonlinear filtering methods
	\cite{Davis,schutz1,cutlerflan94,cardiff} that could be used
	to give better results for parameter measurements (see
	Ref.~\cite{note2} above). The use of such   
	methods would be more computationally intensive than linear filtering,
	but on the other hand, vast amounts of supercomputer time would
	have already been expended to generate merger templates.
\bibitem{noteinfo} 
	No significant changes would result from changing
	the requirement ``have the same magnitude as the measured signal''
	to the requirement ``have a magnitude which is smaller than or equal
	to that of the measured signal''; the numerical value of
	the information would be approximately the same.  The reason
	we insert this requirement into our definition of information
	[so that we use the probability distribution $p[{\bf h} \, | \,
	\rho({\bf s}) ]$ in Eq.~(\ref{infodef}) rather than
	the probability distribution $p^{(0)}({\bf h})$] is that the
	resulting notion of information is more useful and relevant 
	than the notion obtained from using $p^{(0)}({\bf h})$.  If
	one were to use the PDF $p^{(0)}$, the resulting 
	``information gain in the measurement'' would be the same for
	all detected signals, whether the SNR be 10 or 1000.
\bibitem{noteprior}
	The PDF $p^{(0)}({\bf h})$ describes all one's prior
	information about the signal ${\bf h}$, except for the information that
	the maximum expected duration is $T$ and that the maximum expected
	bandwidth is $\Delta f$; that information is already encoded in the
	choice of the space $V$.
\bibitem{notedetect1} Note that the regime $\rho \gg {\cal N}_{\rm
	bins}^{1/4}$ coincides with the regime in which the signal is
	detectable using the noise-monitoring search method discussed in
	Refs.~\cite{paperI,ComingSoonToAJournalNearYou}.
\bibitem{noteI} In fact, this result is independent of the specific
	form of the norm on the space of pairs of waveforms $\{ h_+(t),
	h_\times(t) \}$ chosen in Eq.~(\ref{GTinnerprod}).
\bibitem{caveat1} In Eqs.~(\ref{prior00}), (\ref{identity}),
	(\ref{pp2}), (\ref{pp3}), and (\ref{pp5}) 
	we omit factors of ${\cal N}-2$ dimensional solid angle elements
	$d \Omega^{{\cal N}-2} / C_{{\cal N}-2}$ which integrate to one.
	Thus, the equals symbol in, for example, Eq.~(\ref{prior00}) should be
	interpreted to mean that the PDFs on the left-hand-side and
	right-hand-side will produce  
	identical expected values of functions that depend on $\theta$
	and $h$ only.
\end{references}
\end{document}